\documentclass[aps,prd,twocolumn,eqsecnum]{revtex4-2}

\begin{document}

\title{Komar superpotentials in extended theories of  gravity}

\author{H. Arthur Weldon}
\email[]{hweldon@mail.wvu.edu}
\affiliation{Department of Physics and Astronomy, West Virginia University, 
Morgantown, West Virginia, 26506, USA}

\date{September 10, 2026}

\begin{abstract}

This paper investigates Lagrangians with arbitrary dependence on the Riemann tensor and
any number of  derivatives. Noether's second theorem is applied separately to ${\cal L}_{G}$ and ${\cal L}_{M}$,
where the matter Lagrangian depends on a  scalar field. For ${\cal L}_{G}$ diffeomorphism invariance
requires that the generalized Einstein tensor satisfy $\nabla_{\alpha}E^{\alpha\beta}=0$ and that there is a current satisfying $0=\partial_{\alpha}(\sqrt{g}\, J^{\alpha}_{G2})$.  The main result is an explicit formula for the
superpotential $\Phi^{[\alpha\mu]}$ that solves $\partial_{\mu}\Phi^{[\alpha\mu]}=\sqrt{g}\,J^{\alpha}_{G2}$.
Some applications to $f(R)$ gravity
and various quadratic gravity theories are discussed.  There are analogous results, including superpotentials,
for the matter sector whenever the couplings to the matter fields contain derivatives of the metric.

\end{abstract}

\maketitle

\section{Introduction \label{Intro}}

The consequences of diffeomorphism invariance for
any gravitational action
are expressed in Noether's second theorem 
\cite{Noether,Utiyama,DeHaro,McLeod,Ilin1,Ilin2,Aoki,eSa1,eSa2,Freese}.
Invariance  under an
infinitesimal coordinate transformation
\begin{equation}
x^{\lambda}\to x^{\lambda}+\xi^{\lambda}(x^{0},x^{1},x^{2},x^{3})\label{100}\end{equation}
where $\xi^{\lambda}$ has arbitrary  spacetime dependence
leads to an identity of the form
\begin{equation}
0={\cal E}+\partial_{\alpha}{\cal J}^{\alpha}\label{102}\end{equation}
where both terms are linear in $\xi$ and its derivatives. This identity, due to Noether,
expresses general diffeomorphism invariance and holds for any values of the metric and the matter fields.

Historically this principle has been applied  to the total Lagrangian for gravitation and matter with
 ${\cal L}_{G}=\sqrt{g}\,{\rm R}/16\pi G$, where $g=-{\rm det}(g_{\mu\nu})>0$. The results are often presented as follows. 
 First, if the field equations for the metric are satisfied then  ${\cal E}=0$.
 Second, the current has the form
\begin{equation}
{\cal J}^{\alpha}={\cal J}^{\alpha}_{1}+{\cal J}^{\alpha}_{2}\label{104}\end{equation}
where ${\cal J}^{\alpha}_{1}$ vanishes when ${\cal E}$ vanishes, i.e. when the field equations are satisfied.
Third, 
 $\partial_{\alpha}{\cal J}^{\alpha}_{2}=0$ for any metric without regard
 for the field equations. And finally, the conserved current can be obtained from a superpotential
 ${\cal J}^{\alpha}_{2}=\partial_{\mu}\Phi^{[\alpha\mu]}$ with
\begin{equation}
\Phi^{[\alpha\mu]}={\sqrt{g}\over 16\pi G}(\nabla^{\mu}\xi^{\alpha}-\nabla^{\alpha}\xi^{\mu})\label{105}\end{equation}
These  results apparently led Einstein and many others to define the energy-momentum of the gravitational
field by  pseudotensors \cite{Einstein,Moller,Dirac,Landau}.

In 1959 Komar  used various physical arguments, not Noether's theorem, to argue that the volume integral
\begin{equation}
Q_{\xi}= \int_{\cal V} d^{3}x\, \partial_{j}\Phi^{[0j]}=\int_{\partial \cal V}dS_{j}\Phi^{[0j]}
 \end{equation}
 is a sensible accounting of the energy, momentum, and angular momentum of the gravitational field for appropriate
 choices of $\xi$ \cite{Komar1,Komar2,Komar3,Townsend}.   Thus $\Phi^{[\alpha\mu]}$ is called the Komar superpotential even though
 it was previously known from  Noether's second theorem.

There are now  extended theories gravity  \cite{Capozziello1,Odintsov1,Clifton,Odintsov2}  motivated by cosmological problems (dark energy and inflation, dark matter and large scale structure) and also by quantum gravity.
One class includes Lagrangians of the form $\sqrt{g}\, f(R)$, where $R$ is the Ricci scalar \cite{Capozziello2,DeFelice1,Sotiriou,Kumar,Montani}.
Another category are  Lagrangians that are linear combinations of the scalars $R^{2}$,
$R_{\alpha\beta}R^{\alpha\beta}$, and $R_{\rho\sigma\tau\omega}R^{\rho\sigma\tau\omega}$
\cite{Alvarez-Gaume,Holdom,Salvio,Daas,Donoghue,Brito,Buccio,Kuntz}.
A special type of the quadratic form is Gauss-Bonnet gravity where the quadratic term 
is a total derivative \cite{DeFelice2,Glavan,Fernandes}.

In  extended theories of gravity it is known that Noether's second theorem
\cite{Ilin1,Ilin2,eSa1,eSa2}
leads to conserved currents that depend on an arbitrary vector $\xi$.  
The main task of this paper is to find explicit formulas for the corresponding superpotentials.

\paragraph*{Outline and summary:}
Sections II, III, and IV investigate a general class of extended theories  with a Lagrangian of the form
\begin{equation}
{\cal L}={\cal L}_{G}(g_{\ast\ast},{\rm R}_{\ast\ast\ast\ast})
+{\cal L}_{M}(g_{\ast\ast},R_{\ast\ast\ast\ast},\phi,\partial_{\mu}\phi)\label{121}\end{equation}
where matter is represented by the spin zero field $\phi$. 
Both terms can depend on   scalars formed by 
contractions of the metric with  Riemann tensors. Lagrangians of this type 
include $f(R)$ theories and  quadratic theories and allows
more complicated terms
such as ${\rm R}^{\rho\tau}{\rm R}^{\sigma\omega}{\rm R}_{\rho\sigma\tau\omega}$,
 ${\rm R}^{\rho}_{\;\lambda}{\rm R}^{\lambda\sigma\tau\omega}{\rm R}_{\rho\sigma\tau\omega}$, or ${\rm R}_{\alpha\beta\rho\sigma}{\rm E}^{\rho\sigma\tau\omega}{\rm R}_{\tau\omega\mu\nu}
 g^{\alpha\mu}{\rm R}^{\beta\nu}$ and also couplings to matter such as ${\rm R}\phi^{2}$ or ${\rm R}^{\mu\nu}(\partial_{\mu}\phi)(\partial_{\nu}\phi)$.
 
 It is customary to consider the diffeomorphism invariance of the total Lagrangian, which leads to
 ${\cal E}$ in (\ref{102}) being a linear combination of the field equations, but it is more productive
 to  treat the  diffeomorphism invariance of ${\cal L}_{G}$ and of ${\cal L}_{M}$ separately.

 In Sec. II diffeomorphism invariance is expressed in terms of the Lie derivative.
For ${\cal L}_{G}$ of the type (\ref{121}) Sec. III shows that the generalized Einstein tensor 
 \begin{equation}
 \sqrt{g}\,E^{\alpha\beta}={\partial{\cal L}_{G}\over\partial g_{\alpha\beta}}-\partial_{\mu}
 {\partial{\cal L}_{G}\over\partial(\partial_{\mu}g_{\alpha\beta})}+\partial_{\mu}\partial_{\nu}
 {\partial{\cal L}_{G}\over\partial(\partial_{\mu}\partial_{\nu}g_{\alpha\beta})}\label{130}\end{equation}
 satisfies 
 \begin{equation}\nabla_{\alpha}E^{\alpha\beta}=0\label{131}\end{equation}
  for any metric, without regard to the field
 equations.  
 The Noether identity (\ref{102}) is
 \begin{equation} 0=2\sqrt{g}\,E^{\alpha\beta}\nabla_{\alpha}\xi_{\beta}+\partial_{\alpha}
 \big\{-2\sqrt{g}\,E^{\alpha\beta}\xi_{\beta}+{\cal J}^{\alpha}_{G2}\big\}\label{132} \end{equation}
 and because of (\ref{131}) this simplifies to 
 \begin{equation}\partial_{\alpha}{\cal J}^{\alpha}_{G2}=0.\end{equation}

 This current is  expected to be expressible as the derivative of a superpotential defined by
 $\partial_{\mu}\Phi^{[\alpha\mu]}={\cal J}^{\alpha}_{G2}$.
 Section III displays the formula for the superpotential, which is  proven in Appendix B.1.
  In terms of the tensor $M$
 \begin{equation}
\sqrt{g}\, M^{\alpha\mu\nu\beta}={\partial{\cal L}_{G}\over\partial R_{\alpha\mu\nu\beta}}\label{135}
\end{equation}
the superpotential is
\begin{equation}
\Phi^{[\alpha\mu]}_{G}=4\sqrt{g}\,(\nabla_{\nu}M^{\alpha\mu\nu\beta})\xi_{\beta}
-2\sqrt{g}\,M^{\alpha\mu\nu\beta}\nabla_{\nu}\xi_{\beta},\label{138}
\end{equation}
\Big[ which was found by other means in \cite{Ortin}\Big].
As a by-product the generalized Einstein tensor is expressed as 
\begin{eqnarray}
{\rm E}^{\alpha\beta}&=&2\nabla_{\mu}\nabla_{\nu}M^{\alpha\{\mu\nu\}\beta}
- {1\over 2} M^{\alpha\mu\nu\phi}R^{\beta}_{\;\;\mu\nu\phi}\nonumber\\
&-&{1\over 2} M^{\beta\mu\nu\phi}{\rm R}^{\alpha}_{\;\;\mu\nu\phi}
+{1\over 2}g^{\alpha\beta}{\cal L}_{G}/\sqrt{g}.\label{136}
\end{eqnarray}

Section IV investigates
  matter Lagrangians of the form (\ref{121}) with  energy-momentum tensor 
 \begin{equation}
{1\over 2} \sqrt{g}\,T^{\alpha\beta}={\partial{\cal L}_{M}\over\partial g_{\alpha\beta}}-\partial_{\mu}
 {\partial{\cal L}_{M}\over\partial(\partial_{\mu}g_{\alpha\beta})}+\partial_{\mu}\partial_{\nu}
 {\partial{\cal L}_{M}\over\partial(\partial_{\mu}\partial_{\nu}g_{\alpha\beta})}.\label{140}\end{equation}
 As in the gravitational sector, there is a current ${\cal J}^{\alpha}_{M2}$
satisfying $\partial_{\alpha}{\cal J}^{\alpha}_{M2}=0$ that can be expressed in terms of a superpotential
${\cal J}^{\alpha}_{M2}=\partial_{\mu}\widetilde{\Phi}^{[\alpha\mu]}$. The superpotential has the same form
as (\ref{138}) in terms of a tensor $\widetilde{M}$ analogous  to (\ref{135}). 
The matter analogue of (\ref{136}) is an expression for the energy-momentum tensor 
  \begin{eqnarray}
T^{\alpha\beta}&=&{1\over\sqrt{g}}\Big[-{\partial{\cal L}_{M}\over\partial(\partial_{\alpha}\phi)}
\nabla^{\beta}\phi+g^{\alpha\beta}{\cal L}_{M}\Big]\\
&+&4\nabla_{\mu}\nabla_{\nu}\widetilde{M}^{\alpha\{\mu\nu\}\beta}
-\widetilde{M}^{\alpha\mu\nu\phi}R^{\beta}_{\;\;\mu\nu\phi}-\widetilde{M}^{\beta\mu\nu\phi}R^{\alpha}_{\;\;\mu\nu\phi}\nonumber
\end{eqnarray}
where the first line is  the canonical energy-momentum tensor.
 
 Section V summarizes the results when the metric contains $N$ derivatives of the metric. The proof
 of the general  results are contained in Appendices A and B. 

\paragraph*{Notation:} The  signature of the metric is $(-+++)$, $g=-{\rm det}(g_{\mu\nu})>0$, and   $R^{\alpha}_{\;\;\sigma\tau\omega}=\partial_{\omega} \Gamma^{\alpha}_{\sigma\tau}+\dots$.

\section{Diffeomorphism invariance and  the Lie derivative \label{II.Diff}}

The infinitesimal coordinate transformation (\ref{100}) produces an infinitesimal change in the metric
 and in the scalar field $\phi$. Noether \cite{Noether} denoted the variation in these quantities 
by  $\overline{\delta}$ as have others \cite{Utiyama,Aoki,Goldberg,Bergmann}; however   $\overline{\delta}=-\pounds_{\xi}$, where $\pounds_{\xi}$ is the Lie derivative,
and $\pounds_{\xi}$   will be used here to define infinitesimal variations. This means that the
Noether current and superpotential has  the opposite sign to that used by
early authors. 

The Lie derivative of any  $\Psi(x)$ that changes to $\Psi^{\prime}(x^{\prime})$ under the infinitesimal
coordinate transformation (\ref{100}) is
\begin{equation}
\pounds_{\xi}\Psi(x)=\Psi(x)-\Psi^{\prime}(x^{\prime})+\xi^{\lambda}\partial_{\lambda}\Psi.\label{200}\end{equation}

(a) Both Lagrangians ${\cal L}_{G}$ and ${\cal L}_{M}$ are scalar densities: either can be written as
${\cal L}=\sqrt{g}\,L$ where $L$ is a scalar satisfying 
$L(x)=L^{\prime}(x^{\prime})$. The Lie derivative of $L$ is
\begin{equation}
\pounds_{\xi}L=\xi^{\lambda}\partial_{\lambda}L.\label{201}\end{equation}

(b) The transformation of the metric to first order is
\begin{eqnarray}
g^{\prime}_{\alpha\beta}(x^{\prime})&=&{\partial x^{\mu}\over\partial x^{\prime\alpha}}
{\partial x^{\nu}\over\partial x^{\prime\beta}}g_{\mu\nu}(x)\nonumber\\
&\approx& g_{\alpha\beta}(x) -(\partial_{\alpha}\xi^{\lambda})g_{\lambda\nu}
-(\partial_{\beta}\xi^{\lambda})g_{\alpha\lambda}
\end{eqnarray}
and fixes  the Lie derivative as
\begin{equation}
\pounds_{\xi}g_{\alpha\beta}=\xi^{\lambda}\partial_{\lambda}g_{\alpha\beta}+
(\partial_{\alpha}\xi^{\lambda})g_{\lambda\beta}+(\partial_{\beta}\xi^{\lambda})g_{\alpha\lambda}.\label{203}\end{equation}
This  can also be written in terms of covariant derivatives
\begin{equation}
\pounds_{\xi}g_{\alpha\beta}=\nabla_{\alpha}\xi_{\beta}+\nabla_{\beta}\xi_{\alpha}.\label{204}\end{equation}

(c) Because  ${\cal L}$ is proportional to $\sqrt{g}$ one needs
\begin{equation}
\pounds_{\xi}\sqrt{g}={1\over 2}\sqrt{g}\, g^{\alpha\beta}\pounds_{\xi}g_{\alpha\beta}=\partial_{\lambda}(\xi^{\lambda}
\sqrt{g}\,).\label{205}
\end{equation}
Combining this with (\ref{201}) gives
\begin{equation}
\pounds_{\xi}{\cal L}=\partial_{\lambda}(\xi^{\lambda}{\cal L}).\label{206}\end{equation}
This is the statement of  diffeomorphism invariance that will be used repeatedly.  
It is Eq. (11) in Ref. \cite{Noether} and  will be called the Noether identity.

(d)  Since $\Psi^{\prime}(x^{\prime})-\xi^{\lambda}\partial_{\lambda}\Psi=\Psi^{\prime}(x)$ 
the definition (\ref{200}) can be be stated  as
\begin{equation}
\pounds_{\xi}\Psi(x)=\Psi(x)-\Psi^{\prime}(x).\label{210}\end{equation}
This shows that when  $\Psi$ is the derivative of another function say $\partial_{\mu}\Omega$  then
\begin{equation}
\pounds_{\xi}(\partial_{\mu}\Omega)=\partial_{\mu}\Omega(x)-\partial_{\mu}\Omega^{\prime}(x)=\partial_{\mu}\pounds_{\xi}\Omega, \label{212}\end{equation}
which will be used without comment  in the following.

(e) One can  rearrange (\ref{210}) as
\begin{equation}
\Psi^{\prime}(x)=\Psi(x)-\pounds_{\xi}\Psi(x).\end{equation}
which shows that the Lie derivative can be viewed as an infinitesimal variation in  $\Psi$.
In the following $\Psi$  can be either $g_{\alpha\beta}(x)$,  $\phi(x)$, or their derivatives.

\section{The gravitational Lagrangian  ${\cal L}_{G}$ \label{III.Gravitation}}

This section  investigates   gravitational Lagrangians (\ref{121})  with arbitrary dependence on the Riemann tensor but with no derivatives of the Riemann tensor.
Diffeomorphism invariance (\ref{206}) implies
\begin{equation}
0=\pounds_{\xi}{\cal L}_{G}-\partial_{\alpha}(\xi^{\alpha}{\cal L}_{G}).\label{300}\end{equation}
The Riemann tensor  contains first and second derivatives of the metric and so this becomes
\begin{eqnarray}
0&=&{\partial{\cal L}_{G}\over\partial g_{\alpha\beta}}\pounds_{\xi}g_{\alpha\beta}
+{\partial{\cal L}_{G}\over\partial(\partial_{\mu}g_{\alpha\beta})}\partial_{\mu}\pounds_{\xi}
g_{\alpha\beta}\nonumber\\
&& +{\partial{\cal L}_{G}\over\partial(\partial_{\mu}\partial_{\nu} g_{\alpha\beta})}
\partial_{\mu}\partial_{\nu}\pounds_{\xi}g_{\alpha\beta}-\partial_{\alpha}(\xi^{\alpha}{\cal L}_{G}).\label{301}
\end{eqnarray}
It is useful to  reorganize this as
\begin{equation}
0=\sqrt{g}\,{\rm E}^{\alpha\beta}\pounds_{\xi}g_{\alpha\beta}+\partial_{\alpha}{\cal J}^{\alpha}_{G},\label{303}\end{equation}
where  $E^{\alpha\beta}$ is given in  (\ref{130}) and the Noether current is
\begin{eqnarray}
{\cal J}^{\alpha}_{G}&=&\bigg[{\partial{\cal L}_{G}\over\partial(\partial_{\alpha}g_{\rho\sigma})}
-\partial_{\nu}{\partial{\cal L}_{G}\over\partial(\partial_{\alpha}\partial_{\nu}g_{\rho\sigma})}\bigg]\pounds_{\xi}g_{\rho\sigma}\nonumber\\
&&+{\partial{\cal L}_{G}\over\partial(\partial_{\alpha}\partial_{\nu}g_{\rho\sigma})}\partial_{\nu}\pounds_{\xi}g_{\rho\sigma}
-\xi^{\alpha}{\cal L}_{G}.\label{304}\end{eqnarray}
The current  has terms with second-order derivatives of $\xi$, first-order derivatives of $\xi$, and no derivatives of $\xi$ and can be expanded as
\begin{equation}
J^{\alpha}_{G}=A^{\alpha}_{\lambda}\xi^{\lambda}+B_{\lambda}^{\alpha\mu}\partial_{\mu}\xi^{\lambda}
+B_{\lambda}^{\alpha\{\mu_{1},\mu_{2}\}}\partial_{\mu_{1}}\partial_{\mu_{2}} \xi^{\lambda}\label{306}\end{equation}
This type of derivative expansion is standard \cite{Utiyama,Aoki,Ilin1,Ilin2,eSa2}.
The coefficients can be expressed as functional derivatives of ${\cal L}_{G}$ but the explicit forms
will not be needed until Sec. III.B.1  and Appendix C.1. The current will turn out to be a vector density;
see (\ref{341}).

 \subsection{The road to the superpotential}
 
\subsubsection{Proof that $\nabla_{\alpha}E^{\alpha}_{\;\;\lambda}=0$} 
Because  (\ref{303}) holds for any $\xi$, substitution of  the derivative expansion (\ref{306})  into
(\ref{303})    results is four independent equations: 
 \begin{eqnarray}
 0&=&B_{\lambda}^{\alpha\{\mu_{1}\mu_{2}\}}\partial_{\alpha}\partial_{\mu_{1}}\partial_{\mu_{2}}\xi^{\lambda}
\label{308a}\\
0&=&\Big\{B_{\lambda}^{\mu_{1}\mu_{2}}+\partial_{\alpha}B^{\alpha\{\mu_{1}\mu_{2}\}}_{\lambda}\Big\}\partial_{\mu_{1}}\partial_{\mu_{2}}\xi^{\lambda}\label{308b}\\
0&=&\Big\{\sqrt{g}\,2 E^{\alpha}_{\;\;\lambda}
+A^{\alpha}_{\lambda}+\partial_{\mu}B^{\mu\alpha}_{\lambda}\Big\}\partial_{\alpha}\xi^{\lambda}\label{308c}\\
0&=&\Big\{\sqrt{g}\,E^{\alpha\beta}\partial_{\lambda}g_{\alpha\beta}+\partial_{\alpha}A^{\alpha}_{\lambda}\Big\}\xi^{\lambda}.\label{308d}
\end{eqnarray}
The first equation requires that the fully symmetrized, three-superscript coefficient vanish 
\begin{equation}
B_{\lambda}^{\{\alpha\mu_{1}\mu_{2}\}}=0.\label{311a}\end{equation}
The second equation (\ref{308b}) requires that the fully symmetrized, two-superscript coefficient  satisfy
\begin{equation}
0= B^{\{\mu_{1}\mu_{2}\}}_{\lambda}+\partial_{\alpha}B_{\lambda}^{\alpha\{\mu_{1}\mu_{2}\}}.\label{311b}\end{equation}
Application of $\partial_{\mu_{1}}\partial_{\mu_{2}}$ to this gives, because of (\ref{311a}), 
\begin{equation}
0=\partial_{\mu_{1}}\partial_{\mu_{2}}B^{\{\mu_{1}\mu_{2}\}}_{\lambda}.\label{311c}\end{equation}
The third equation (\ref{308c})  requires
\begin{equation}
A^{\alpha}_{\lambda}=-\sqrt{g}\,2 E^{\alpha}_{\;\;\lambda}
-\partial_{\mu}B^{\mu\alpha}_{\lambda}.\label{311d}\end{equation}
Application of $\partial_{\alpha}$ to this gives, because of  (\ref{311c}),
\begin{equation}
\partial_{\alpha}A^{\alpha}_{\lambda}=-\partial_{\alpha}\Big[\sqrt{g}\,2E^{\alpha}_{\;\;\lambda}\Big].
\label{311e}
\end{equation}
Substitution  into the fourth equation (\ref{308d}) gives
\begin{eqnarray}0&=&E^{\alpha\beta}\partial_{\lambda}g_{\alpha\beta}
-\partial_{\alpha}\Big[\sqrt{g}\,2E^{\alpha}_{\;\;\lambda}\Big]\nonumber\\
&=&-2\sqrt{g}\,\nabla_{\alpha}E^{\alpha}_{\;\;\lambda}.\label{311f}
\end{eqnarray}
 This is true for any metric without imposing the  field equations.

\subsubsection{The conserved current ${\cal J}^{\alpha}_{G2}$\label{DerivativeExpansion}}
Substitution of (\ref{311d}) into the derivative expansion (\ref{306}) suggests splitting ${\cal J}^{\alpha}_{G}$
into two parts
\begin{equation}
{\cal J}^{\alpha}_{G}={\cal J}^{\alpha}_{G1}+{\cal J}^{\alpha}_{G2},\end{equation}
where 
\begin{eqnarray}
{\cal J}^{\alpha}_{G1}&=&-2\sqrt{g}\,E^{\alpha}_{\;\;\lambda}\xi^{\lambda}\label{318}\\
{\cal J}^{\alpha}_{G2}&=&-(\partial_{\mu}B_{\lambda}^{\mu\alpha})\xi^{\lambda}+B^{\alpha\mu}_{\lambda}\partial_{\mu}\xi^{\lambda}
+B_{\lambda}^{\alpha\{\mu_{1}\mu_{2}\}}\partial_{\mu_{1}}\partial_{\mu_{2}}\xi^{\lambda}.\nonumber
\end{eqnarray}
The Noether identity (\ref{303}) is now
\begin{eqnarray}
0&=&2\sqrt{g}\,E^{\alpha\beta}\nabla_{\alpha}\xi_{\beta}+\partial_{\alpha}\Big[ -2\sqrt{g}\, E^{\alpha}_{\;\;\lambda}\xi^{\lambda}+{\cal J}^{\alpha}_{G2}\Big]\nonumber\\
&=&-2\sqrt{g}\,(\nabla_{\alpha}(E^{\alpha}_{\;\;\lambda})\xi^{\lambda}
+\partial_{\alpha}{\cal J}^{\alpha}_{G2}\nonumber\\
&=&\partial_{\alpha}{\cal J}^{\alpha}_{G2}\label{320}
\end{eqnarray}
because of (\ref{311f}).
The result  holds for any metric without using the field equations. This type of conservation law
was termed ``improper" by Noether \cite{Noether}.

\paragraph*{Another  proof:} A further check is to differentiate the derivative expansion
of ${\cal J}^{\alpha}_{G2}$  in (\ref{318}):
\begin{eqnarray}
\partial_{\alpha}{\cal J}^{\alpha}_{G2}&=&-(\partial_{\alpha}\partial_{\mu}B_{\lambda}^{\mu\alpha})\xi^{\lambda}\nonumber\\
&&-(\partial_{\mu}B^{\mu\alpha}_{\lambda})(\partial_{\alpha}\xi^{\lambda})
+(\partial_{\alpha}B^{\alpha\mu}_{\lambda})(\partial_{\mu}\xi^{\lambda})\nonumber\\
&&+B_{\lambda}^{\alpha\mu}\partial_{\alpha}\partial_{\mu}\xi^{\lambda}+(\partial_{\alpha}B_{\lambda}^{\alpha\{\mu_{1}\mu_{2}\}})
\partial_{\mu_{1}}\partial_{\mu_{2}}\xi^{\lambda}\nonumber\\
&&+B_{\lambda}^{\alpha\{\mu_{1}\mu_{2}\}}\partial_{\alpha}\partial_{\mu_{1}}\partial_{\mu_{2}}\xi^{\lambda}.
\label{322}\end{eqnarray}
The first term vanishes because of (\ref{311c}); the two terms on the second line cancel automatically;
the two terms on the third line cancel because of (\ref{311b}); the fourth line vanishes because of  (\ref{311a}).

\paragraph*{Comment:} Though different physically, a simple example of a superpotential is
the field strength tensor $\sqrt{g}\,F^{\alpha\mu}$ of electrodynamics. The electromagnetic current
${\cal J}^{\alpha}_{em}=\partial_{\mu}(\sqrt{g}\, F^{\alpha\mu})$  is conserved because of the antisymmetry of
$F^{\alpha\mu}$ and ${\cal J}^{\alpha}_{em}$  can be expressed in terms of the
vector potential: $A^{\lambda}$, $\partial_{\mu}A^{\lambda}$, and 
$\partial_{\mu_{1}}\partial_{\mu_{2}}A^{\lambda}$ as in (\ref{318}). 

\subsubsection{Formula for the superpotential}
 The superpotential satisfying $\partial_{\mu}\Phi^{[\alpha\mu]}={\cal J}^{\alpha}_{G2}$ is
 \begin{eqnarray}
&&\Phi^{[\alpha\mu_{1}]}=\Big[B_{\lambda}^{[\alpha\mu_{1}]}\!-\!{1\over 3}(\partial_{\mu_{2}}C_{\lambda}^{[\alpha\mu_{1}]\mu_{2}})
\Big]\xi^{\lambda}+{2\over 3}C_{\lambda}^{[\alpha\mu_{1}]\mu_{2}}\partial_{\mu_{2}}\xi^{\lambda}\nonumber\\
&&C_{\lambda}^{[\alpha\mu_{1}]\mu_{2}}=B_{\lambda}^{\alpha\{\mu_{1}\mu_{2}\}}
-B_{\lambda}^{\mu_{1}\{\alpha\mu_{2}\}}.\label{324}
\end{eqnarray}
 The proof is given in  Appendix B.1. 

\subsection{Manifestly covariant results  \label{Cov}}
From the previous formulas for  ${\cal J}^{\alpha}_{G2}$ and $\Phi^{[\alpha\mu]}$
it is unclear how they behave under general coordinate transformations and whether they
can be easily computed from the Lagrangian.
This section  addresses  those matters.

The coefficients in the derivative expansion of the current (\ref{306}) are  determined by the dependence of the Lagrangian on the  first and second
derivatives of the metric. The Lagrangian under consideration depends on the Riemann tensor but not on its derivatives and so  derivatives of the metric occur only from the dependence
of the Lagrangian on the Riemann tensor. It  is natural to define
\begin{equation}
{\partial{\cal L}_{G}\over\partial R_{\alpha\mu\nu\beta}}=\sqrt{g}\,M^{\alpha\mu\nu\beta}.\label{338}\end{equation}
The tensor  $M$ has the properties
\begin{eqnarray}
&&M^{\alpha\mu\nu\beta}=-M^{\mu\alpha\nu\beta}=-M^{\alpha\mu\beta\nu}={\rm M}^{\nu\beta\alpha\mu}
\label{340}\\
&&0=M^{\alpha\mu\nu\beta}+M^{\alpha\nu\beta\mu}+M^{\alpha\beta\mu\nu}.\label{340b}
\end{eqnarray}
The Noether current (\ref{304}) requires the derivative
\begin{eqnarray}
{\partial{\cal L}_{G}\over\partial(\partial_{\alpha}\partial_{\beta}g_{\mu\nu})}
&=&{\partial{\cal L}_{G}\over\partial R_{\rho\sigma\tau\omega}}{\partial R_{\rho\sigma\tau\omega}
\over\partial(\partial_{\alpha}\partial_{\beta}g_{\mu\nu})}\nonumber\\
&=&\sqrt{g}\,\big[M^{\alpha\mu\nu\beta}+M^{\alpha\nu\mu\beta}\big]\\
&=& 2\sqrt{g}\,M^{\alpha\{\mu\nu\}\beta}.\nonumber
\end{eqnarray}
Note that $M^{\mu\{\alpha\beta\}\nu}=M^{\alpha\{\mu\nu\}\beta}$ and thus is symmetric in both
$\alpha\beta$ and $\mu\nu$.
The dependence  of ${\cal L}_{G}$ on the first derivative of the metric is also due to the Riemann tensor:
\begin{eqnarray}
{\partial{\cal L}_{G}\over\partial(\partial_{\alpha}g_{\mu\nu})}&=&{\partial{\cal L}_{G}\over\partial {\rm R}_{\rho\sigma\tau\omega}}{\partial{\rm R}_{\rho\sigma\tau\omega}\over\partial(\partial_{\alpha}g_{\mu\nu})}\nonumber\\
&=&2\sqrt{g}\,\Big[\Gamma^{\mu}_{\rho\omega}M^{\rho\{\alpha\nu\}\omega}+\Gamma^{\nu}_{\rho\omega}M^{\rho\{\alpha\mu\}\omega}\nonumber\\
&&\hskip1cm -\Gamma^{\alpha}_{\rho\omega}M^{\rho\{\mu\nu\}\omega}\Big].
\end{eqnarray}

\subsubsection{Covariant  Noether current}
The   Noether current  is determined by the  tensor $M$.
The remaining obstacle is to convert the ordinary derivatives $\partial_{\nu}$ in (\ref{304}) into covariant derivatives
$\nabla_{\nu}$. This is performed in Appendix  C.1 with the result
\begin{eqnarray}
{\cal J}^{\alpha}_{G}&=&\sqrt{g}\,\Big[-2(\nabla_{\beta}M^{\alpha\{\mu\nu\}\beta})\pounds_{\xi}g_{\mu\nu}\nonumber\\
&+&4M^{\alpha\{\beta\nu\}\mu}\nabla_{\mu}\nabla_{\nu}\xi_{\beta}
-\xi^{\alpha}L_{G}\Big].\label{341}\end{eqnarray}

\subsubsection{Covariant superpotential}
The next step is to express  ${\cal J}^{\alpha}_{G}$ in terms of ordinary derivatives of $\xi$ as in (\ref{306}), identify the coefficients
$B^{\alpha\mu}_{\lambda}$ and $B^{\alpha\{\mu\nu\}}_{\lambda}$, and substitute into  formula
(\ref{324}) for the superpotential. These steps are performed in Appendix C.2 and yield \cite{Ortin}
\begin{equation}
\Phi^{[\alpha\mu]}=4\sqrt{g}(\nabla_{\nu}M^{\alpha\mu\nu\beta})\xi_{\beta} -2\sqrt{g}\,M^{\alpha\mu\nu\beta}\nabla_{\nu}\xi_{\beta}.\label{343a}
\end{equation}
This can be  expressed in the more general form
\begin{eqnarray}
\Phi^{[\alpha\mu]}&=&\sqrt{g}\,\Big[-r\nabla_{\nu}(M^{\alpha\mu\nu\beta}\xi_{\beta})\label{343b}\\
&+&(r+4)(\nabla_{\nu}M^{\alpha\mu\nu\beta})\xi_{\beta}
+(r-2)M^{\alpha\mu\nu\beta}\nabla_{\nu}\xi_{\beta}\Big]\nonumber
\end{eqnarray}
for any real parameter $r$.

\subsubsection{Covariant form for  ${\cal J}^{\alpha}_{G2}$}
The divergence of the superpotential is
\begin{eqnarray}
{\cal J}^{\alpha}_{G2}&=&\partial_{\mu}\Phi^{[\alpha\mu]}\nonumber\\
&=&\sqrt{g}\,\Big[4(\nabla_{\mu}\nabla_{\nu}M^{\alpha\mu\nu\beta})\xi_{\beta}
-2(\nabla_{\beta}M^{\alpha\{\mu\nu\}\beta})\pounds_{\xi}g_{\mu\nu}\nonumber\\
&&-2M^{\alpha\mu\nu\beta}\nabla_{\mu}\nabla_{\nu}\xi_{\beta}\Big]\label{345}
\end{eqnarray}
independent of $r$.
As expected it has first and second derivatives of $\xi$.

\subsubsection{Covariant form of the  generalized Einstein tensor} 
The covariant form for ${\cal J}^{\alpha}_{G1}$ is obtained by subtraction:
\begin{equation}
{\cal J}^{\alpha}_{G1}={\cal J}^{\alpha}_{G}-{\cal J}^{\alpha}_{G2}.\end{equation}
The terms proportional to $\pounds_{\xi}g_{\alpha\beta}$ in (\ref{341}) and   (\ref{345})  cancel.
Subtraction of the terms containing $\nabla_{\mu}\nabla_{\nu}\xi_{\beta}$ gives
\begin{eqnarray}
&&\sqrt{g}\,\Big[4M^{\alpha\{\beta\nu\}\mu}+2M^{\alpha\mu\nu\beta}\Big]\nabla_{\mu}\nabla_{\nu}\xi_{\beta}\nonumber\\
&&\hskip1cm =-2\sqrt{g}\,M^{\alpha\beta\mu\nu}(\nabla_{\mu}\nabla_{\nu}-\nabla_{\nu}\nabla_{\mu})\xi_{\beta}\\
&&\hskip1cm=2\sqrt{g}\,M^{\alpha\mu\nu\phi}R^{\beta}_{\;\;\mu\nu\phi}\xi_{\beta}.\nonumber
\end{eqnarray}
The difference between (\ref{341}) and (\ref{345}) is therefore
\begin{equation}
{\cal J}^{\alpha}_{G1}=-2\sqrt{g}\,{\rm S}^{\alpha\beta}\xi_{\beta},\label{350}\end{equation}
where the coefficient is
\begin{equation}
{\rm S}^{\alpha\beta}=2\nabla_{\mu}\nabla_{\nu}M^{\alpha\mu\nu\beta}- M^{\alpha\mu\nu\phi}{\rm R}^{\beta}_{\;\;\mu\nu\phi}+{1\over 2}g^{\alpha\beta}L_{G}\label{351}.
\end{equation}
Despite its appearance, $S^{\alpha\beta}$ is symmetric and has zero covariant divergence.

\paragraph*{Divergence of $S^{\alpha\beta}$:} To compute the covariant divergence of the first term in
(\ref{351})
use the $\alpha\mu$ antisymmetry of $M^{\alpha\mu\nu\beta}$ to obtain
\begin{eqnarray}
\nabla_{\alpha} S^{\alpha\beta}&=&(\nabla_{\alpha}\nabla_{\mu}-\nabla_{\mu}\nabla_{\alpha})
\nabla_{\nu}M^{\alpha\mu\nu\beta}\nonumber\\
&-&(\nabla_{\alpha}M^{\alpha\mu\nu\phi})R^{\beta}_{\;\;\mu\nu\phi}\\
&-&M^{\alpha\mu\nu\phi}\nabla_{\alpha}R^{\beta}_{\;\;\mu\nu\phi}+{1\over 2}\nabla^{\beta}L_{G}.\nonumber
\end{eqnarray}
The commutator of the covariant derivatives gives a Riemann tensor that cancels the second line.
Only the last line remains and it is convenient to lower the $\beta$ index and to again use the
$\alpha\mu$ antisymmetry of $M^{\alpha\mu\nu\beta}$:
\begin{eqnarray}
\nabla_{\alpha}S^{\alpha}_{\;\;\beta}&=&-{1\over 2}M^{\alpha\mu\nu\phi}
\big(\nabla_{\alpha}R_{\beta\mu\nu\phi}+\nabla_{\mu}R_{\alpha\beta\nu\phi}\big)
+{1\over 2}\nabla_{\beta}L_{G}\nonumber\\
&=&-{1\over 2}M^{\alpha\mu\nu\phi}\nabla_{\beta}R_{\alpha\mu\nu\phi}+{1\over 2}\nabla_{\beta}L_{G}
\end{eqnarray}
after using the Bianchi identity. In view of the definition of $M$ this is
\begin{equation}
\nabla_{\alpha}S^{\alpha}_{\;\;\beta}=-{1\over 2}{\partial L_{G}\over\partial R_{\alpha\mu\nu\phi}}
\nabla_{\beta}R_{\alpha\mu\nu\phi}+{1\over 2}\nabla_{\beta}L_{G}.
\end{equation}
Application of $\nabla_{\beta}$ to $L_{G}$ cancels the first term and 
shows that $\nabla_{\alpha}S^{\alpha}_{\;\;\beta}=0$.

\paragraph*{Symmetry of $S^{\alpha\beta}$:}
The antisymmetric part of  ${\rm S}^{\alpha\beta}$ is 
\begin{eqnarray}
{\rm S}^{\alpha\beta}-{\rm S}^{\beta\alpha}&=&2(\nabla_{\mu}\nabla_{\nu}-\nabla_{\nu}\nabla_{\mu})
M^{\alpha\mu\nu\beta}\nonumber\\
&-&M^{\alpha\mu\nu\phi}{\rm R}^{\beta}_{\;\;\mu\nu\phi}+M^{\beta\mu\nu\phi}
{\rm R}^{\alpha}_{\;\;\mu\nu\phi}.
\end{eqnarray}
This vanishes because the commutator of the covariant derivatives cancels the second line. 

\paragraph*{Relation to $E^{\alpha\beta}$:}
Using the manifestly symmetric form of $S^{\alpha\beta}$ the current 
(\ref{350}) is
\begin{eqnarray}
{\cal J}^{\alpha}_{G1}&=&-\sqrt{g}\,({\rm S}^{\alpha\beta}+{\rm S}^{\beta\alpha})\xi_{\beta}.\end{eqnarray}
From   (\ref{318})  ${\cal J}^{\alpha}_{G1}=-2\sqrt{g}\,E^{\alpha\beta}\xi_{\beta}$ and so the generalized Einstein tensor is given by
$E^{\alpha\beta}=(S^{\alpha\beta}+S^{\beta\alpha})/2$.  The explicitly symmetric form is
\begin{eqnarray}
{\rm E}^{\alpha\beta}&=&2\nabla_{\mu}\nabla_{\nu}M^{\alpha\{\mu\nu\}\beta}
- {1\over 2} M^{\alpha\mu\nu\phi}R^{\beta}_{\;\;\mu\nu\phi}\nonumber\\
&-&{1\over 2} M^{\beta\mu\nu\phi}{\rm R}^{\alpha}_{\;\;\mu\nu\phi}
+{1\over 2}g^{\alpha\beta}L_{G}.\label{352}
\end{eqnarray}

\subsection{Examples}
\subsubsection{ $f({\rm R})$ gravity} 
The Lagrangian ${\cal  L}_{G}=\sqrt{g}\,f({\rm R})/16\pi G$ has attractive features for cosmology
\cite{Odintsov1,Odintsov2,Capozziello2,DeFelice1,Sotiriou,Kumar,Montani}.
The tensor ${\rm M}$ is
\begin{equation}{\rm M}^{\alpha\mu\nu\beta}=
{f^{\prime}({\rm R})\over 32\pi G}(g^{\alpha\nu}g^{\beta\mu}-g^{\alpha\beta}g^{\mu\nu})
\end{equation}
which gives a  superpotential \cite{Ortin}
\begin{eqnarray}\Phi^{[\alpha\mu]}&=&{\sqrt{g}\over 16\pi G}\Big[f^{\prime}(R)(\nabla^{\mu}\xi^{\alpha}
-\nabla^{\alpha}\xi^{\mu})\nonumber\\
&+&2f^{\prime\prime}(R)\big((\partial^{\alpha}R)\xi^{\mu}-(\partial^{\mu} R)
\xi^{\alpha}\big)\Big].\end{eqnarray}

\subsubsection{Quadratic gravity}
The Lagrangian for quadratic gravity  \cite{Alvarez-Gaume,Holdom,Salvio,Daas,Donoghue,Brito,Buccio,Kuntz}
is
\begin{equation}
{\cal L}_{G}=\sqrt{g}\,\Big[aR^{2}+bR^{\mu\nu}R_{\mu\nu}+cR^{\rho\sigma\tau\omega}R_{\rho\sigma\tau\omega}\Big]\label{360}\end{equation}
where $a,b,c$ are parameters that distinguish various sub-cases.
The superpotential (\ref{343a}) requires  the tensor $M$ 
\begin{eqnarray}M^{\alpha\mu\nu\beta}&=&a\big(g^{\alpha\nu}g^{\beta\mu}-g^{\alpha\beta}g^{\mu\nu}\big)R\nonumber\\
&&\hskip-0.5cm +{b\over 2}\big(g^{\alpha\nu}R^{\beta\mu}+g^{\beta\mu}R^{\alpha\nu}
-g^{\alpha\beta}R^{\mu\nu}-g^{\mu\nu}R^{\alpha\beta}\big)\nonumber\\
&+&2cR^{\alpha\mu\nu\beta}\label{362}
\end{eqnarray}
and its covariant derivative
\begin{eqnarray}
\nabla_{\nu}M^{\alpha\mu\nu\beta}&=&\big(a+{b\over 4}\big)\big(g^{\beta\mu}\nabla^{\alpha}R-g^{\alpha\beta}\nabla^{\mu}R\big)\label{364}\\
&+&{b\over 2}\big(\nabla^{\alpha}R^{\beta\mu}-\nabla^{\mu}R^{\alpha\beta}\big)
+2c\nabla_{\nu}R^{\alpha\mu\nu\beta}\nonumber.\end{eqnarray}
The Einstein tensor is
\begin{eqnarray}
E^{\alpha\beta}&=&(a+{b\over 2}+c)(\nabla^{\alpha}\nabla^{\beta}+\nabla^{\beta}\nabla^{\alpha})R\nonumber\\
&-&(2a+{b\over 2})g^{\alpha\beta}\nabla_{\mu}\nabla^{\mu}R
-(b+4c)\nabla_{\mu}\nabla^{\mu}R^{\alpha\beta}\nonumber\\
&+&-2aRR^{\alpha\beta}
+4cR^{\alpha}_{\;\;\lambda}R^{\lambda\beta}\label{366}\\
&+&2(b+2c)R_{\mu\nu}R^{\alpha\{\mu\nu\}\beta}-2cR^{\alpha\mu\nu\lambda}R^{\beta}_{\;\;\mu\nu\lambda}\nonumber\\
&+&{g^{\alpha\beta}\over 2}(aR^{2}+bR_{\mu\nu}R^{\mu\nu}+cR^{\rho\sigma\tau\omega}
R_{\rho\sigma\tau\omega}).\nonumber
\end{eqnarray}

\paragraph{Conformal gravity:} A special case of (\ref{360}) is conformal gravity. The  Lagrangian  is
\begin{equation}{\cal L}_{G}=c\sqrt{g}\,C^{\rho\sigma\tau\omega}C_{\rho\sigma\tau\omega},\end{equation}
where $C$ is the Weyl tensor
\begin{eqnarray}
C_{\rho\sigma\tau\omega}&=&{1\over 6}\Big(g_{\rho\tau}g_{\sigma\omega}-g_{\rho\omega}g_{\sigma\tau}\Big)R
+R_{\rho\sigma\tau\omega}\\
&&\hskip-1cm -{1\over 2}\Big(g_{\rho\tau}R_{\sigma\omega}+g_{\sigma\omega}R_{\rho\tau}
-g_{\rho\omega}R_{\sigma\tau}-g_{\sigma\tau}R_{\rho\omega}\Big).\nonumber
\end{eqnarray}

The parameters in  (\ref{360}) are
\begin{equation}
a={1\over 3}c;\hskip0.5cm b=-2c\end{equation}
Neither the superpotential nor the Einstein tensor are particularly simple.

\paragraph{Gauss-Bonnet gravity:} For the values 
\begin{equation}
a=c;\hskip0.5cm b=-4c\end{equation} 
the second-order derivatives of $R$ and $R^{\alpha\beta}$ in (\ref{366}) are absent and 
\begin{eqnarray}
E^{\alpha\beta}&=&c\Big[-2RR^{\alpha\beta}+4R^{\alpha}_{\;\;\lambda}R^{\beta\lambda}\nonumber\\
&-&4\,R_{\mu\nu}R^{\alpha\{\mu\nu\}\beta}
-2R^{\alpha\mu\nu\lambda}R^{\beta}_{\;\;\mu\nu\lambda}\label{375}\\
&+&{1\over 2}g^{\alpha\beta}\big(R^{2}-4R_{\mu\nu}R^{\mu\nu}
+R^{\rho\sigma\tau\omega}R_{\rho\sigma\tau\omega}\big)\Big]\nonumber
\end{eqnarray}
This is Gauss-Bonnet gravity \cite{Odintsov1,Odintsov2,DeFelice2,Glavan, Fernandes}.  
The Lagrangian is a total derivative in four dimensions and so the Einstein tensor should vanish.
As expected  (\ref{375})  does vanish due to  a known identity in four dimensions \cite{Glavan,Fernandes}.
This is a stringent check of  (\ref{366}) and  of the basic formula (\ref{352}).

Even though the Gauss-Bonnet term is a total derivative,  the tensor $M$ is not zero:
\begin{eqnarray}
M^{\alpha\mu\nu\beta}&=&c\Big[(g^{\alpha\nu}g^{\beta\mu}-g^{\alpha\beta}g^{\mu\nu})R+2R^{\mu\alpha\beta\nu}\\
&-&2(g^{\alpha\nu}R^{\beta\mu}+g^{\beta\mu}R^{\alpha\nu}-g^{\alpha\beta}R^{\mu\nu}
-g^{\mu\nu}R^{\alpha\beta})\Big].\nonumber
\end{eqnarray}
and the superpotential (\ref{343a}) is non-zero. This,  because the Noether identity (\ref{300}) is a property
of the Lagrangian not of the action.

\section{The matter Lagrangian  ${\cal L}_{M}$ \label{IV.Matter}}

Analysis similar to Sec. III  will be applied to a matter Lagrangian with arbitrary dependence on the Riemann tensor
and on  a scalar fields $\phi$:
\begin{equation}
{\cal L}_{M}(g_{\alpha\beta}, R_{\rho\sigma\tau\omega}, \phi,\partial_{\mu}\phi).\label{400}\end{equation}
Matter Lagrangians with no Riemann tensor but with second-order derivatives of the scalar field will contain first-order derivatives of the metric from the Christoffel connection
and are treated in Sec. IV.D.

\subsection{Diffeomorphism invariance  of ${\cal L}_{M}$}
Diffeomorphism  invariance of (\ref{400}) requires
\begin{equation}
0=\pounds_{\xi}{\cal L}_{M}-\partial_{\alpha}(\xi^{\alpha}{\cal L}_{M})\label{405}\end{equation}
or more explicitly
\begin{eqnarray}
0&=&{\partial{\cal L}_{M}\over\partial\phi}\pounds_{\xi}\phi
+{\partial{\cal L}_{M}\over\partial(\partial_{\mu}\phi)}\partial_{\mu} \pounds_{\xi}\phi
-\partial_{\alpha}(\xi^{\alpha}{\cal L}_{M})\nonumber\\
&+&{\partial{\cal L}_{M}\over\partial g_{\alpha\beta}}\pounds_{\xi}g_{\alpha\beta}
+{\partial{\cal L}_{M}\over\partial(\partial_{\mu}g_{\alpha\beta})}\partial_{\mu}\pounds_{\xi}g_{\alpha\beta}
\nonumber\\
&+&{\partial{\cal L}_{M}\over\partial(\partial_{\mu}\partial_{\nu}g_{\alpha\beta})}\partial_{\mu}\partial_{\nu}
\pounds_{\xi}g_{\alpha\beta}.\end{eqnarray}
This can be organized as
\begin{eqnarray}
0&=&{1\over 2}\sqrt{g}\,{\rm T}^{\alpha\beta}\pounds_{\xi}g_{\alpha\beta}+\Big[{\partial{\cal L}_{M}\over\partial\phi}
-\partial_{\mu}{\partial{\cal L}_{M}\over\partial(\partial_{\mu}\phi)}\Big]\pounds_{\xi}\phi\nonumber\\
&&+\partial_{\alpha}{\cal J}^{\alpha}_{M},\label{406}
\end{eqnarray}
where the energy-momentum tensor of matter is given in  ({\ref{140}) and the Noether current is
\begin{eqnarray}
{\cal J}^{\alpha}_{M}&=&{\partial{\cal L}_{M}\over\partial(\partial_{\alpha}\phi)}\pounds_{\xi}\phi
-\xi^{\alpha}{\cal L}_{M}\nonumber\\
&+&\Big[{\partial{\cal L}_{M}\over\partial(\partial_{\alpha}g_{\rho\sigma})}-\partial_{\nu}
{\partial{\cal L}_{M}\over\partial(\partial_{\nu}\partial_{\alpha}g_{\rho\sigma})}\Big]\pounds_{\xi}g_{\rho\sigma}\nonumber\\
&+&{\partial{\cal L}_{M}\over\partial(\partial_{\alpha}\partial_{\nu}g_{\rho\sigma})}
\partial_{\nu}\pounds_{\xi}g_{\rho\sigma}
\label{407a}.\end{eqnarray}
 The first line is $-\sqrt{g}\,\Theta^{\alpha}_{\;\;\lambda}\xi^{\lambda}$ 
where $\Theta^{\alpha}_{\;\;\lambda}$ is the canonical energy-momentum tensor.

\subsection{Analysis of ${\cal J}^{\alpha}_{M}$}
\subsubsection{Derivative expansion of  ${\cal J}^{\alpha}_{M}$}
As in III.A  the next step is to parametrize the current in terms of derivatives of $\xi$:
\begin{equation}
J^{\alpha}_{M}=\widetilde{A}^{\alpha}_{\lambda}\xi^{\lambda}+\widetilde{B}_{\lambda}^{\alpha\mu}\partial_{\mu}\xi^{\lambda}
+\widetilde{B}_{\lambda}^{\alpha\{\mu_{1},\mu_{2}\}}\partial_{\mu_{1}}\partial_{\mu_{2}} \xi^{\lambda}\label{408}.\end{equation}
The expansion coefficients are analogous to those in (\ref{306})  but  depend on 
both the metric and the field $\phi$. 

Substitution of (\ref{408}) into the Noether identity (\ref{406}) gives three equations involving derivatives of $\xi$
\begin{eqnarray}
0&=&\widetilde{B}_{\lambda}^{\alpha\{\mu_{1}\mu_{2}\}}\partial_{\alpha}\partial_{\mu_{1}}\partial_{\mu_{2}}\xi^{\lambda}
\label{409a}\\
0&=&\Big\{\widetilde{B}_{\lambda}^{\mu_{1}\mu_{2}}+\partial_{\alpha}\widetilde{B}^{\alpha\{\mu_{1}\mu_{2}\}}_{\lambda}\Big\}\partial_{\mu_{1}}\partial_{\mu_{2}}\xi^{\lambda}\label{409b}\\
0&=&\Big\{\sqrt{g}\,T^{\alpha}_{\;\;\lambda}
+\widetilde{A}^{\alpha}_{\lambda}+\partial_{\mu}\widetilde{B}^{\mu\alpha}_{\lambda}\Big\}\partial_{\alpha}\xi^{\lambda}
\label{409c}\end{eqnarray}
and a fourth equation with no derivatives of $\xi$
\begin{eqnarray}
&&0=\bigg\{{1\over 2}\sqrt{g}\,T^{\alpha\beta}\partial_{\lambda}g_{\alpha\beta}\label{409d}\nonumber\\
&&\hskip0.2cm +\Big[{\partial{\cal L}_{M}\over\partial\phi}
-\partial_{\mu}{\partial{\cal L}_{M}\over\partial(\partial_{\mu}\phi)}\Big]\partial_{\lambda}\phi
+\partial_{\alpha}\widetilde{A}^{\alpha}_{\lambda}\bigg\}\xi^{\lambda}.\end{eqnarray}
The first equation (\ref{409a}) requires 
\begin{equation}
\widetilde{B}_{\lambda}^{\{\alpha\mu_{1}\mu_{2}\}}=0.\label{411}\end{equation}
The second  (\ref{409b}) requires 
\begin{equation}
0= \widetilde{B}^{\{\mu_{1}\mu_{2}\}}_{\lambda}+\partial_{\alpha}\widetilde{B}_{\lambda}^{\alpha\{\mu_{1}\mu_{2}\}},\label{412}\end{equation}
which implies
\begin{equation}
0=\partial_{\mu_{1}}\partial_{\mu_{2}}\widetilde{B}^{\{\mu_{1}\mu_{2}\}}_{\lambda}.\label{413}\end{equation}
The third equation (\ref{409c})  requires
\begin{equation}
\widetilde{A}^{\alpha}_{\lambda}=-\sqrt{g}\,T^{\alpha}_{\;\;\lambda}
-\partial_{\mu}\widetilde{B}^{\mu\alpha}_{\lambda},\label{414}\end{equation}
which leads to
\begin{equation}
\partial_{\alpha}\widetilde{A}^{\alpha}_{\lambda}=-\partial_{\alpha}\Big[\sqrt{g}\,T^{\alpha}_{\;\;\lambda}\Big].
\end{equation}
Substitution of (\ref{414})  into  (\ref{409d}) gives
\begin{equation}
0=-\sqrt{g}\,\nabla_{\alpha}T^{\alpha}_{\;\;\lambda}+
\Big[{\partial{\cal L}_{M}\over\partial\phi}-\partial_{\mu}{\partial{\cal L}_{M}\over\partial(\partial_{\mu}\phi)}\Big]
\partial_{\lambda}\phi\label{415}
\end{equation}
This relation is the analogue of $\nabla_{\alpha}E^{\alpha}_{\;\;\lambda}=0$; it 
  holds for any metric and any scalar  field  $\phi$.
Only when the $\phi$ satisfies the Euler-Lagrange equation does the
energy-momentum tensor satisfy covariant conservation.

\subsubsection{The conserved current ${\cal J}^{\mu}_{M2}$}}
Substitution of (\ref{414})  into the derivative expansion (\ref{408}) suggests splitting ${\cal J}^{\alpha}_{M}$
into two parts
\begin{equation}
{\cal J}^{\alpha}_{M}={\cal J}^{\alpha}_{M1}+{\cal J}^{\alpha}_{M2},\end{equation}
where 
\begin{eqnarray}
{\cal J}^{\alpha}_{M1}&=&-\sqrt{g}\,T^{\alpha}_{\;\;\lambda}\xi^{\lambda}\label{417}\\
{\cal J}^{\alpha}_{M2}&=&-(\partial_{\mu}\widetilde{B}^{\mu\alpha})\xi^{\lambda}+\widetilde{B}^{\alpha\mu}_{\lambda}\partial_{\mu}\xi^{\lambda}
+\widetilde{B}_{\lambda}^{\alpha\{\mu_{1}\mu_{2}\}}\partial_{\mu_{1}}\partial_{\mu_{2}}\xi^{\lambda}
.\nonumber\end{eqnarray}
The Noether identity (\ref{406}) is now
\begin{eqnarray}
0&=&\bigg\{-\sqrt{g}\,\nabla_{\alpha}T^{\alpha}_{\;\;\lambda}+\Big[{\partial{\cal L}_{M}\over\partial\phi}
-\partial_{\mu}{\partial{\cal L}_{M}\over\partial(\partial_{\mu}\phi)}\Big]\partial_{\lambda}\phi\bigg\}
\xi^{\lambda}\nonumber\\
&+&\partial_{\alpha}{\cal J}^{\alpha}_{M2}.\nonumber
\end{eqnarray}
The quantity in curly braces always vanishes as shown in (\ref{415}) and therefore
\begin{equation}
0=\partial_{\alpha}{\cal J}^{\alpha}_{M2}\end{equation}
One can also  compute  the divergence from (\ref{417}).

\subsection{Covariant form}
Since ${\cal J}^{\alpha}_{M2}$ is determined by $\widetilde{B}^{\alpha\mu}_{\lambda}$ and
$\widetilde{B}^{\alpha\{\mu_{1}\mu_{2}\}}_{\lambda}$ 
it can be expressed in terms of the tensor
\begin{equation}
\sqrt{g}\, \widetilde{M}^{\alpha\mu\nu\beta}={\partial{\cal L}_{M}\over\partial R_{\alpha\mu\nu\beta}}
\label{420}\end{equation}
analogous to  (\ref{338}) except for its dependence on $\phi$.
The conserved current is
\begin{eqnarray}
{\cal J}^{\alpha}_{M2}
&=&\sqrt{g}\,\Big[4(\nabla_{\mu}\nabla_{\nu}\widetilde{M}^{\alpha\mu\nu\beta})\xi_{\beta}
-2(\nabla_{\beta}\widetilde{M}^{\alpha\{\mu\nu\}\beta})\pounds_{\xi}g_{\mu\nu}\nonumber\\
&&-2\widetilde{M}^{\alpha\mu\nu\beta}\nabla_{\mu}\nabla_{\nu}\xi_{\beta}\Big].
\end{eqnarray}
The superpotential satisfying $\partial_{\mu}\widetilde{\Phi}^{[\alpha\mu]}={\cal J}^{\alpha}_{M2}$ is
\begin{equation}
\widetilde{\Phi}^{[\alpha\mu]}=4\sqrt{g}\,(\nabla_{\nu}\widetilde{M}^{\alpha\mu\nu\beta})\xi_{\beta}
-2\sqrt{g}\,\widetilde{M}^{\alpha\mu\nu\beta}\nabla_{\nu}\xi_{\beta}.
\end{equation}
The dependence of the energy-momentum tensor on the Riemann tensor is
\begin{eqnarray}
T^{\alpha\beta}&=&{1\over\sqrt{g}}\Big[-{\partial{\cal L}_{M}\over\partial(\partial_{\alpha}\phi)}
\nabla^{\beta}\phi+g^{\alpha\beta}{\cal L}_{M}\Big]\label{460}\\
&+&4\nabla_{\mu}\nabla_{\nu}\widetilde{M}^{\alpha\{\mu\nu\}\beta}
-\widetilde{M}^{\alpha\mu\nu\phi}R^{\beta}_{\;\;\mu\nu\phi}-\widetilde{M}^{\beta\mu\nu\phi}R^{\alpha}_{\;\;\mu\nu\phi}.\nonumber
\end{eqnarray}
The first line is the canonical energy-momentum tensor.

\subsubsection*{Example}
An natural possibility is to allow the  scalar field to couple to the Ricci scalar:
\begin{equation}{\cal L}_{M}={\cal L}_{0}(g_{\alpha\beta}, \phi,\partial_{\mu}\phi)+
\sqrt{g}\,\gamma\phi^{2}R\end{equation}
where $\gamma$ is a real coupling constant.
The tensor $\widetilde{M}$ is
\begin{equation}
\widetilde{M}^{\alpha\mu\nu\beta}={\gamma\over 2}\phi^{2}(g^{\alpha\nu}g^{\beta\mu}
-g^{\alpha\beta}g^{\mu\nu}),\end{equation}
and the  superpotential is
\begin{eqnarray}
\widetilde{\Phi}^{[\alpha\mu]}&=&\sqrt{g}\,\gamma\phi^{2}(\nabla^{\mu}\xi^{\alpha}-\nabla^{\alpha}\xi^{\mu})
\nonumber\\
&+& 4\sqrt{g}\,\gamma\phi(\xi^{\mu}\nabla^{\alpha}\phi-\xi^{\alpha}\nabla^{\mu}\phi).
\end{eqnarray}
The first line is essentially the original Komar form (\ref{105}).
It is  simpler to compute the energy-momentum tensor from (\ref{460}) than from the
definition (\ref{140}):
\begin{eqnarray}
T^{\alpha\beta}&=&{1\over\sqrt{g}}\Big[-{\partial{\cal L}_{0}\over\partial(\partial_{\alpha}\phi)}
\nabla^{\beta}\phi +g^{\alpha\beta}{\cal L}_{0}\Big]\\
&+&\gamma\phi^{2}(g^{\alpha\beta}R-2R^{\alpha\beta})\nonumber\\
&+&4\gamma\Big[(\nabla^{\alpha}\phi)(\nabla^{\beta}\phi)-g^{\alpha\beta}
(\nabla_{\mu}\phi)(\nabla^{\mu}\phi)\Big]\nonumber\\
&+&4\gamma\phi\Big[{1\over 2}(\nabla^{\alpha}\nabla^{\beta}+\nabla^{\beta}\nabla^{\alpha})
\phi-g^{\alpha\beta}\nabla_{\mu}\nabla^{\mu}\phi\Big].\nonumber
\end{eqnarray}

\subsection{Matter Lagrangians with no Riemann tensor}
The previous discussion in Sec. IV  treated matter Lagrangians in which first-order and second-order derivatives of the metric occur through the Riemann tensor. In the first category below the Lagrangian contains only first-order
derivatives of the metric but it results in a  conserved current with a superpotential.
The second category below has no derivatives of the metric and contains no conserved current. 

\subsubsection{Matter Lagrangians  second-order derivatives of  $\phi$ but no Riemann tensor  }
If one allows second-order derivatives of the scalar field
$\nabla_{\{\mu}\nabla_{\nu\}}\phi$ then the necessary Christoffel connection contains first-order derivatives
of the metric and these will lead to a superpotential even in the absence of the Riemann tensor. Thus consider
\begin{equation}
{\cal L}_{M}(g_{\alpha\beta},\phi,\nabla_{\mu}\phi,\nabla_{\{\mu}\nabla_{\nu\}}\phi).
\end{equation}
Diffeomorphism invariance leads to
\begin{eqnarray}
0&=&{1\over 2}\sqrt{g}\,T^{\alpha\beta}\pounds_{\xi}g_{\alpha\beta}\nonumber\\
&+&\Big[{\partial{\cal L}_{M}\over\partial\phi}-\partial_{\mu}{\partial{\cal L}_{M}\over\partial
(\partial_{\mu}\phi)}+\partial_{\mu}\partial_{\nu}{{\cal L}_{M}\over\partial(\partial_{\mu}\partial_{\nu}
\phi)}\Big]\pounds_{\xi}\phi\nonumber\\
&+&\partial_{\alpha}{\cal J}^{\alpha}_{M}.\label{470}
\end{eqnarray}
The Noether current ${\cal J}^{\alpha}_{M}$ is simpler than (\ref{407a}) since now ${\cal L}_{M}$ has no
second-order derivatives of the metric. The derivative expansion of ${\cal J}^{\alpha}_{M}$ is simpler
than (\ref{408}):  $\widetilde{B}_{\lambda}^{\alpha\{\mu_{1}\mu_{2}\}}=0$ and $\widetilde{B}^{\{\alpha\mu\}}_{\lambda}=0$;
thus $\widetilde{B}^{\alpha\mu}_{\lambda}=\widetilde{B}_{\lambda}^{[\alpha\mu]}$.
The energy-momentum tensor satisfies
\begin{eqnarray}
0&=&-\sqrt{g}\,\nabla_{\alpha}T^{\alpha}_{\lambda}\\
&+&\Big[{\partial{\cal L}_{M}\over\partial\phi}-\partial_{\mu}{\partial{\cal L}_{M}\over\partial
(\partial_{\mu}\phi)}+\partial_{\mu}\partial_{\nu}{{\cal L}_{M}\over\partial(\partial_{\mu}\partial_{\nu}
\phi)}\Big]\partial_{\lambda}\phi\nonumber\end{eqnarray}
The Noether identity (\ref{470}) reduces  as usual to current conservation
$0=\partial_{\alpha}{\cal J}^{\alpha}_{M2}$ where now the superpotential form is automatic:
\begin{equation}
{\cal J}^{\alpha}_{M2}=-(\partial_{\mu}\widetilde{B}^{\mu\alpha}_{\lambda})\xi^{\lambda}
+\widetilde{B}^{\alpha\mu}_{\lambda}\partial_{\mu}\xi^{\lambda}
=\partial_{\mu}(\widetilde{B}_{\lambda}^{[\alpha\mu]}).
\end{equation}
In terms of functional derivatives of the Lagrangian
\begin{equation}
\widetilde{B}_{\lambda}^{[\alpha\mu]}=g_{\lambda\sigma}\Big[{\partial{\cal L}_{M}\over\partial(\partial_{\alpha}g_{\mu\sigma})}-{\partial{\cal L}_{M}\over\partial(\partial_{\mu}g_{\alpha\sigma})}\Big]
\end{equation}
Since the derivatives of the metric come entirely from the Christoffel connection  in $\nabla_{\{\mu}\nabla_{\nu\}}\phi$
it is natural to define
\begin{equation}\sqrt{g}\,M^{\alpha\beta}={\partial{\cal L}_{M}\over\partial(\nabla_{\{\alpha}\nabla_{\beta\}}
\phi)}\end{equation}
which allows the superpotential to be written covariantly:
\begin{equation}
\Phi^{[\alpha\mu]}=\sqrt{g}\,\Big[(\nabla^{\alpha}\phi)M^{\mu}_{\;\;\lambda}
-(\nabla^{\mu}\phi)M^{\alpha}_{\;\;\lambda}\Big]\xi^{\lambda}.\end{equation}

\subsubsection{Mater Lagrangians with no derivatives of the metric and no Riemann tensor}
The simplest matter Lagrangian has no Riemann tensor and only first-order derivatives of the
scalar field:
\begin{equation}
{\cal L}_{M}(g_{\alpha\beta}, \phi,\nabla_{\mu}\phi).\end{equation}
Diffeomorphism invariance gives
\begin{eqnarray}
0&=&{1\over 2}\sqrt{g}\, T^{\alpha\beta}\pounds_{\xi}g_{\alpha\beta}\\
&+&\Big[ {\partial{\cal L}_{M}\over\partial{\phi}}-\partial_{\mu}{\partial{\cal L}_{M}\over\partial(\partial_{\mu}\phi)}\Big]\pounds_{\xi}\phi+\partial_{\alpha}{\cal J}^{\alpha}_{M}.\nonumber
\end{eqnarray}
The Noether current is
\begin{equation}
{\cal J}^{\alpha}_{M}=\Big[{\partial{\cal L}_{M}\over\partial(\partial_{\alpha}\phi)}\pounds_{\xi}\phi
-\delta^{\alpha}_{\lambda}{\cal L}_{M}\Big]\xi^{\lambda}.
\end{equation}
One recovers the previous definition of the energy-momentum tensor  (\ref{460}) but without the tensor $\widetilde{M}$
and the previous formula for the divergence of the energy-momentum tensor (\ref{415}).  There is no
conserved current and consequently no superpotential.

\section{Discussion}
 Appendices \ref{Current} and \ref{Superpotential} generalize the analysis  to  Lagrangians with any  number of derivatives of the Riemann tensor 
 such as $g^{\mu\nu}(\partial_{\mu}{\rm R})(\partial_{\nu}{\rm R})$ or $(\nabla^{\mu}{\rm R}^{\alpha\beta})(\nabla_{\mu}{\rm R}_{\alpha\beta})$ 
 and more generally with $N-2$ derivatives of the Riemann tensor \cite{Wald,Biswas}:
   \begin{equation}
 {\cal L}_{G}(g_{\ast\ast}, R_{\ast\ast\ast\ast}, \nabla_{\mu_{1}}R_{\ast\ast\ast\ast},
 .. \nabla_{\{\mu_{1}}\nabla_{\mu_{2}}..\nabla_{\mu_{N-2}\}}R_{\ast\ast\ast\ast})\label{500}
 \end{equation}
 and thus $N$ derivatives of the metric. 
 As before,  applying Noether's identity
 to ${\cal L}_{G}$ and ${\cal L}_{M}$ separately  produces more results than if applied to the sum.
 
 For the gravitational Lagrangian Appendix A shows that the generalized Einstein tensor 
\begin{equation}
\sqrt{g}\,E^{\alpha\beta}=\sum_{s=0}^{N}(-1)^{s}\partial_{\mu_{1}}..\partial_{\mu_{s}}
{\partial{\cal L}_{G}\over\partial(\partial_{\mu_{1}}..\partial_{\mu_{s}}g_{\alpha\beta})}\label{510}
\end{equation}
satisfies $\nabla_{\alpha}E^{\alpha\beta}=0$
  for any value of the metric. 
  
 Appendix B examines the derivative expansion of the current ${\cal J}^{\alpha}_{G2}$:
 \begin{equation}
 {\cal J}^{\alpha}_{G2}=B^{\alpha}_{\lambda}\xi^{\lambda}+
 \sum_{s=1}^{N}B_{\lambda}^{\alpha\{\mu_{1}..\mu_{s}\}}
 \partial_{\mu_{1}}..\partial_{\mu_{s}}\xi^{\lambda}\label{515}\end{equation} 
  where the coefficients are related to the Lagrangian by expanding (\ref{A102b}) in terms of derivatives of $\xi$. 
The  superpotential  antisymmetric in $\alpha\mu_{1}$ must have the expansion
\begin{equation}
\Phi^{[\alpha\mu_{1}]}=W_{\lambda}^{[\alpha\mu_{1}]}\xi^{\lambda}
+\sum_{s=2}^{N}W_{\lambda}^{[\alpha\mu_{1}]\mu_{2}..\mu_{s}}
\partial_{\mu_{2}}..\partial_{\mu_{s}}\xi^{\lambda}.\label{535}
\end{equation}
Appendix B.4 shows that  $\partial_{\mu_{1}}\Phi^{[\alpha\mu_{1}]}={\cal J}^{\alpha}_{G2}$ is solved by the $W$'s given in Eq. (\ref{B411}).

\begin{appendix}

\section{The conserved current when   ${\cal L}_{G}$ contains $N$ derivatives of the metric \label{Current}}

This Appendix and the next one treat Lagrangians with  $N-2$  covariant derivatives of the Riemann tensor and thus $N$ derivatives of the metric. Either ${\cal L}_{G}$ or ${\cal L}_{M}$ could contain $N$ derivatives of the
metric. Here  ${\cal L}_{G}$ will be analyzed:
 \begin{displaymath}
 {\cal L}_{G}(g_{\ast\ast}, R_{\ast\ast\ast\ast}, \nabla_{\mu_{1}}R_{\ast\ast\ast\ast},
 \dots \nabla_{\{\mu_{1}}\nabla_{\mu_{2}}\dots\nabla_{\mu_{N-2}\}}R_{\ast\ast\ast\ast}).
 \end{displaymath}
 Applying the results  to ${\cal L}_{M}$ is straightforward. 
 Appendix A shows the following: (1)  the generalized Einstein tensor $E^{\alpha\beta}$
defined in (\ref{510}) satisfies $\nabla_{\alpha}E^{\alpha\beta}=0$ and (2)   there is a conserved current satisfying
$\partial_{\alpha}{\cal J}^{\alpha}_{G2}=0$.  Neither result requires the metric to satisfy the field equations. 

Diffeomorphism invariance requires
\begin{equation}
0=\pounds_{\xi}{\cal L}_{G}-\partial_{\mu}(\xi^{\mu}{\cal L}_{G}).\label{A100}\end{equation}
The more explicit statement is 
\begin{eqnarray}
0&=&{\partial{\cal L}_{G}\over\partial g_{\alpha\beta}}\pounds_{\xi}g_{\alpha\beta}+\sum_{s=1}^{N}{\partial{\cal L}_{G}\over\partial(\partial_{\mu_{1}}
\dots\partial_{\mu_{s}} g_{\alpha\beta})} \partial_{\mu_{1}}
\dots\partial_{\mu_{s}}\pounds_{\xi}g_{\alpha\beta}\nonumber\\
&&-\partial_{\mu}(\xi^{\mu}{\cal L}_{G}),\label{A101}\end{eqnarray}
which  can be reorganized as
\begin{equation}
0=\sqrt{g}\,E^{\alpha\beta}\pounds_{\xi}g_{\alpha\beta}+\partial_{\alpha}{\cal J}^{\alpha}_{G}
\label{A102}.\end{equation}
The Noether current is
\begin{eqnarray}
{\cal J}^{\alpha}_{G}&=&\sum_{\ell=0}^{N}{\cal J}_{\ell}^{\alpha}-\xi^{\alpha}{\cal L}_{G}\label{A102a}\\
{\cal J}^{\alpha}_{\ell}&=&\sum_{p=0}^{\ell} (-1)^{p} \partial_{\mu_{1}}..\partial_{\mu_{p}}
{\partial{\cal L}_{G}\over\partial(\partial_{\alpha}\partial_{\mu_{1}}..\partial_{\mu_{\ell}}g_{\rho\sigma})}\nonumber\\
&&\hskip1cm\times \partial_{\mu_{p+1}}..\partial_{\mu_{\ell}}\pounds_{\xi}g_{\rho\sigma}\label{A102b}
\end{eqnarray}

\subsection{Proof that $\nabla_{\alpha}E^{\alpha\beta}=0$}
The highest derivative of $\xi$ in (\ref{A102b})   comes from  the order $N$ derivative of $\pounds_{\xi}g_{\alpha\beta}$ and so the current has a  derivative expansion
\begin{equation}
{\cal J}^{\alpha}_{G}=A^{\alpha}_{\lambda}\xi^{\lambda}+\sum_{s=1}^{N}B_{\lambda}^{\alpha\{\mu_{1}
\dots\mu_{s}\}}\partial_{\mu_{1}}\dots\partial_{\mu_{s}}\xi^{\lambda}\label{A103}.\end{equation}
The coefficients $A^{\alpha}_{\lambda}$ and $B_{\lambda}^{\alpha\{\mu_{1}..\mu_{s}\}}$
can be expressed as functional derivatives of ${\cal L}_{G}$ 
using (\ref{A102b}).

Since $\xi$ is arbitrary,
substitution of this into (\ref{A102}) gives $N+2$ independent equations:
\begin{eqnarray}
0&=&B_{\lambda}^{\alpha\{\mu_{N}\dots\mu_{1}\}}\partial_{\alpha}\partial_{\mu_{N}}\dots\partial_{\mu_{1}}\xi^{\lambda}\label{A105a}\\
0&=&\Big[B_{\lambda}^{\mu_{N}\{\mu_{N-1}..\mu_{1}\}}\!+\!\partial_{\alpha}B_{\lambda}^{\alpha\{\mu_{N}..\mu_{1}\}}
\Big]\partial_{\mu_{N}}..\partial_{\mu_{1}}\xi^{\lambda}\label{A105b}\\
\vdots\nonumber\\
0&=&\Big[B_{\lambda}^{\mu_{3}\{\mu_{2}\mu_{1}\}}+\partial_{\alpha}B_{\lambda}^{\alpha\{\mu_{3}\mu_{2}\mu_{1}\}}\Big]
\partial_{\mu_{3}}\partial_{\mu_{2}}\partial_{\mu_{1}}\xi^{\lambda}\label{A105c}\\
0&=&\Big[B_{\lambda}^{\mu_{2}\mu_{1}}+\partial_{\alpha}B_{\lambda}^{\alpha\{\mu_{2}\mu_{1}\}}\Big]
\partial_{\mu_{2}}\partial_{\mu_{1}}\xi^{\lambda}\label{A105d}\\
0&=&\Big[2\sqrt{g}\, E^{\alpha}_{\;\;\lambda}+A^{\alpha}_{\;\;\lambda}+\partial_{\mu}B_{\lambda}^{\mu\alpha}\Big]\partial_{\alpha}\xi^{\lambda}\label{A105e}\\
0&=&\Big[\sqrt{g}\,E^{\alpha\beta}\partial_{\lambda}g_{\alpha\beta}+\partial_{\alpha}A^{\alpha}_{\lambda}\Big]\xi^{\lambda}.
\label{A105f}\end{eqnarray}
The first equation (\ref{A105a}) requires the fully symmetrized $B$ with $N+1$ superscripts to vanish:
\begin{equation}
0=B_{\lambda}^{\{\alpha\mu_{N}\dots\mu_{1}\}}.\label{A107a}\end{equation}
The second equation (\ref{A105b}) requires
\begin{equation}
0=B_{\lambda}^{\{\mu_{N}\dots\mu_{1}\}}+\partial_{\alpha}B^{\alpha\{\mu_{N}\dots\mu_{1}\} }.\label{A107b} \end{equation}
Applying $N$ partial derivatives  and using (\ref{A107a}) gives
\begin{equation}
0=\partial_{\mu_{N}}\dots\partial_{\mu_{1}}B_{\lambda}^{\{\mu_{N}\dots\mu_{1}\}}.\label{A107c}\end{equation}
The next equation (not shown above) is
\begin{equation}
0=B_{\lambda}^{\{\mu_{N-1}\dots\mu_{1}\}}+\partial_{\alpha}B^{\alpha\{\mu_{N-1}\dots\mu_{1}\} }\label{A107d}.\end{equation}
Applying $N-1$ partial derivatives  and using (\ref{A107c}) gives
\begin{equation}
0=\partial_{\mu_{N-1}}\dots\partial_{\mu_{1}}B_{\lambda}^{\{\mu_{N-1}\dots\mu_{1}\}}.\label{A107e}\end{equation}
The general result for $2\le s\le N$ is
\begin{eqnarray}
0&=&B_{\lambda}^{\{\mu_{s}\dots\mu_{1}\}}+\partial_{\alpha}B^{\alpha\{\mu_{s}\dots\mu_{1}\} }\label{A107f}\\
0&=&\partial_{\mu_{s}}\dots\partial_{\mu_{1}}B_{\lambda}^{\{\mu_{s}\dots\mu_{1}\}}.\label{A107g}
\end{eqnarray}
The last two equations  (\ref{A105e}) and (\ref{A105f}) are different. 
The first of these requires
\begin{equation}
A^{\alpha}_{\lambda}=-2\sqrt{g}\,E^{\alpha}_{\;\;\lambda}-\partial_{\mu}B^{\mu\alpha}.\label{A110}\end{equation}
Application of $\partial_{\alpha}$ and using (\ref{A107g}) with $s=2$ gives 
\begin{equation}
\partial_{\alpha}A^{\alpha}_{\lambda}=-2\partial_{\alpha}(\sqrt{g}\,E^{\alpha}_{\;\;\lambda}).\end{equation}
Substitution into (\ref{A105f}) gives
\begin{equation}
0=\sqrt{g}\, E^{\alpha\beta}\partial_{\lambda}g_{\alpha\beta}-2\partial_{\alpha}(\sqrt{g}\ E^{\alpha\lambda})\end{equation}
which simplifies to
\begin{equation}
0=-2\sqrt{g}\,\nabla_{\alpha}E^{\alpha}_{\;\;\lambda}.\label{A112}\end{equation}
This holds for any metric.

\subsection{The conserved current $J^{\alpha}_{G2}$}
The formula for $A^{\alpha}_{\lambda}$ in 
 (\ref{A110}) makes it convenient to separate the derivative
  derivative expansion (\ref{A103}) of the current into two parts:
\begin{eqnarray}
&&{\cal J}^{\alpha}_{G1}=-2\sqrt{g}\,E^{\alpha}_{\;\;\lambda}\xi^{\lambda}\\
&&{\cal J}^{\alpha}_{G2}\!=\!-(\partial_{\mu}B_{\lambda}^{\mu\alpha})\xi^{\lambda}
\!+\!\sum_{s=1}^{N}B_{\lambda}^{\alpha\{\mu_{1}
..\mu_{s}\}}\partial_{\mu_{1}}..\partial_{\mu_{s}}\xi^{\lambda}.\hskip0.4cm\label{A117}
\end{eqnarray}
Their sum is 
${\cal J}^{\alpha}_{G}={\cal J}^{\alpha}_{G1}+{\cal J}^{\alpha}_{G2}$.
The Noether identity (\ref{A102}) becomes
\begin{equation}
0=2\sqrt{g}\,E^{\alpha\beta}\nabla_{\alpha}\xi_{\beta}
+\partial_{\alpha}\Big[-2\sqrt{g}\,E^{\alpha\beta}\xi_{\beta}+J^{\alpha}_{G2}\Big],\end{equation}
which simplifies to
\begin{equation}
0=-2\sqrt{g}\,(\nabla_{\alpha}E^{\alpha\beta})\xi_{\beta}+\partial_{\alpha}J^{\alpha}_{G2}.\end{equation}
Because $\nabla_{\alpha}E^{\alpha\beta}=0$ this reduces to
\begin{equation}
0=\partial_{\alpha}J^{\alpha}_{G2}.\label{A120}\end{equation}

\paragraph*{Another proof:} If one applies $\partial_{\alpha}$ to the series expansion  (\ref{A117})
the term with no derivative of $\xi$ is
$-(\partial_{\alpha}\partial_{\mu}B_{\lambda}^{\mu\alpha})$ which vanishes because of (\ref{A107g}). The terms
with one derivative of $\xi$ are
\begin{equation}
-(\partial_{\mu}B_{\lambda}^{\mu\alpha})\partial_{\alpha}\xi^{\lambda}
+(\partial_{\alpha}B_{\lambda}^{\alpha\mu_{1}})\partial_{\mu_{1}}\xi^{\lambda}
\end{equation}
which vanish by inspection. The term in the divergence with $N+1$ derivatives of $\xi$  vanishes
\begin{equation}
B_{\lambda}^{\alpha\{\mu_{1}\dots\mu_{N}\}}\partial_{\alpha}\partial_{\mu_{1}}\dots\partial_{\mu_{N}}\xi^{\lambda}
=0
\end{equation}
because of (\ref{A107a}). 
The terms remaining  are
\begin{eqnarray}
\partial_{\alpha}{\cal J}^{\alpha}_{G2}&=&\sum_{s=1}^{N-1} \Big[B_{\lambda}^{\alpha\{\mu_{1},,\mu_{s}\}}
\partial_{\alpha}\partial_{\mu_{1}}..\partial_{\mu_{s}}\xi^{\lambda}\nonumber\\
&+&(\partial_{\alpha}B_{\lambda}^{\alpha\{\mu_{1},,\mu_{s+1}\}})\partial_{\mu_{1}}..\partial_{\mu_{s+1}}\xi^{\lambda}\Big].\end{eqnarray}
In the first term relabel $\alpha$ as $\mu_{s+1}$. The $s+1$ derivatives  symmetrize the $s+1$ superscripts
and so
\begin{eqnarray}
\partial_{\alpha}{\cal J}^{\alpha}_{G2}&=&\sum_{s=1}^{N-1} \Big[B_{\lambda}^{\{\mu_{1},,\mu_{s+1}\}}
+\partial_{\alpha}B_{\lambda}^{\alpha\{\mu_{1},,\mu_{s+1}\}}\Big]\nonumber\\
&&\times \partial_{\mu_{1}}..\partial_{\mu_{s+1}}\xi^{\lambda}
\end{eqnarray}
This vanishes because of (\ref{A107f}).

\paragraph*{Comment:}  In (\ref{A117}) the coefficient of $\xi^{\lambda}$ is
$-\partial_{\mu}B^{\mu\alpha}_{\lambda}$.  One could begin with
the derivative expansion
\begin{equation}
{\cal J}^{\alpha}_{G2}\!=\!B_{\lambda}^{\alpha}\xi^{\lambda}
\!+\!\sum_{s=1}^{N}B_{\lambda}^{\alpha\{\mu_{1}
..\mu_{s}\}}\partial_{\mu_{1}}..\partial_{\mu_{s}}\xi^{\lambda}.\hskip0.4cm\label{A118}\end{equation}
The condition $\partial_{\alpha}{\cal J}^{\alpha}_{G2}=0$ requires $B^{\alpha}_{\lambda}=-\partial_{\mu_{1}}B^{\mu_{1}\alpha}_{\lambda}$.

\section{The superpotential when ${\cal L}_{G}$ contains $N$ derivatives of the metric\label{Superpotential}}
This Appendix contains the major proofs: B.1 proves the correctness of the $N=2$ superpotential in (\ref{318}); 
B.2 displays the $N=3$ superpotential without proof;  B.3 displays the superpotential for any $N$; and 
B.4 proves the correctness of the  superpotential shown in B.3.

Since ${\cal J}^{\alpha}_{G2}$ in (\ref{A117}) has derivatives of $\xi$ up to order $N$, the superpotential must have
derivatives up to order $N-1$ and have the structure
\begin{equation}
\Phi^{[\alpha\mu_{1}]}=W_{\lambda}^{[\alpha\mu_{1}]}\xi^{\lambda}
\!+\!\sum_{s=2}^{N} W_{\lambda}^{[\alpha\mu_{1}]\mu_{2}\dots\mu_{s}}\partial_{\mu_{2}}\dots\partial_{\mu_{s}}\xi^{\lambda},
\label{B100}\end{equation}
where each  $W$ is  $\alpha\mu_{1}$ antisymmetric.
The derivative  is 
\begin{eqnarray}
&&\partial_{\mu_{1}}\Phi^{[\alpha\mu_{1}]}=(\partial_{\mu_{1}}W_{\lambda}^{[\alpha\mu_{1}]})\xi^{\lambda}
\nonumber\\
&&\hskip0.3cm +\sum_{s=2}^{N}\Big[W_{\lambda}^{[\alpha\mu_{2}]\mu_{3}\dots\mu_{s}}
+\partial_{\mu_{1}}W^{[\alpha\mu_{1}]\mu_{2}..\mu_{s}}\Big]\partial_{\mu_{2}}..\partial_{\mu_{s}}\xi^{\lambda}\nonumber\\
&&\hskip1.3cm +W_{\lambda}^{[\alpha\mu_{1}]\mu_{2}\dots\mu_{n}}\partial_{\mu_{1}}\dots\partial_{\mu_{N}}\xi^{\lambda}.
\label{B101}\end{eqnarray}
The $W$'s must be  such that $\partial_{\mu_{1}}\Phi^{[\alpha\mu_{1}]}$ is
equal to (\ref{A117}). 
The solution for the $W$'s will be expressed in terms of the quantity
\begin{equation}
C_{\lambda}^{[\alpha\mu_{1}]\mu_{2}..\mu_{s}}=B_{\lambda}^{\alpha\{\mu_{1}..\mu_{s}\}}
-B_{\lambda}^{\mu_{1}\{\alpha\mu_{2}..\mu_{s}\}},\label{B102}
\end{equation}
which is antisymmetric in $\alpha\mu_{1}$ and symmetric under all permutations of $\mu_{2}..\mu_{s}$.
Note that for $s=1$
\begin{equation}
{1\over 2}C_{\lambda}^{[\alpha\mu_{1}]}={1\over 2}\Big[B_{\lambda}^{\alpha\mu_{1}}-B_{\lambda}^{\mu_{1}\alpha}\Big]
=B_{\lambda}^{[\alpha\mu_{1}]}.\label{B103}\end{equation}

\subsection{Proof of the  $N=2$ superpotential in (\ref{324}) \label{Proof N=2}}
The main text treated 
 ${\cal L}_{G}$ having no derivatives of the Riemann tensor, i.e. $N=2$.  The superpotential for this case is displayed in (\ref{324}) 
 without proof. The proof will now be given. 
 
 The Noether current ${\cal J}^{\alpha}_{G2}$ in (\ref{318}) contains second-order derivatives of $\xi$;
 the superpotential must contain first-order derivatives:
 \begin{equation}
\Phi^{[\alpha\mu_{1}]}=W_{\lambda}^{[\alpha\mu_{1}]}\xi^{\lambda}
+W_{\lambda}^{[\alpha\mu_{1}]\mu_{2}}\partial_{\mu_{2}}\xi^{\lambda}.
\label{B200}\end{equation}
The divergence is
\begin{eqnarray}
\partial_{\mu_{1}}\Phi^{[\alpha\mu_{1}]}&=&X^{\alpha}_{0}+X_{1}^{\alpha}+X^{\alpha}_{2}\nonumber\\
X^{\alpha}_{0}&=&(\partial_{\mu_{1}}W_{\lambda}^{[\alpha\mu_{1}]})\xi^{\lambda}\label{B202a}\\
X^{\alpha}_{1}&=&\big[W^{[\alpha\mu_{1}]}+\partial_{\mu_{2}}W_{\lambda}^{[\alpha\mu_{2}]\mu_{1}}\big]\partial_{\mu_{1}}\xi^{\lambda}
\label{B202b}\\
X^{\alpha}_{2}&=&W_{\lambda}^{[\alpha\mu_{1}]\mu_{2}}\partial_{\mu_{1}}\partial_{\mu_{2}}\xi^{\lambda}.\label{B202c}
\end{eqnarray}
This section will show that   the solution is
\begin{eqnarray}
W_{\lambda}^{[\alpha\mu_{1}]\mu_{2}}&=&{2\over 3}C_{\lambda}^{[\alpha\mu_{1}]\mu_{2}}\label{B204}\\
W_{\lambda}^{[\alpha\mu_{1}]}&=&{1\over 2}C_{\lambda}^{[\alpha\mu_{1}]}-{1\over 3}\partial_{\mu_{2}}
C_{\lambda}^{[\alpha\mu_{1}]\mu_{2}}.\label{B205}
\end{eqnarray}
The  $C$'s are defined by (\ref{B102}) and the three-superscript $B$ is constrained by
\begin{eqnarray} 0&=&B_{\lambda}^{\{\alpha\mu_{1}\mu_{2}\}}\nonumber\\
&=&{1\over 3}\Big[B_{\lambda}^{\alpha\{\mu_{1}\mu_{2}\}}+B_{\lambda}^{\mu_{1}\{\alpha\mu_{2}\}}
+B_{\lambda}^{\mu_{2}\{\alpha\mu_{1}\}}\Big],\label{B207}\end{eqnarray}
which will be used repeatedly.
The term  with no derivative of $\xi$ is
\begin{eqnarray}
X^{\alpha}_{0}&=&\Big[{1\over 2}\partial_{\mu_{1}}C_{\lambda}^{[\alpha\mu_{1}]}-{1\over 3}\partial_{\mu_{1}}\partial_{\mu_{2}}
C_{\lambda}^{[\alpha\mu_{1}]\mu_{2}}\Big]\xi^{\lambda}\nonumber\\
&=&\Big[\partial_{\mu_{1}}B_{\lambda}^{[\alpha\mu_{1}]}+{2\over 3}\partial_{\mu_{1}}\partial_{\mu_{2}}B_{\lambda}^{\mu_{1}\{\alpha\mu_{2}\}}\nonumber\\
&+&{1\over 3}\partial_{\mu_{1}}\partial_{\mu_{2}}B_{\lambda}^{\mu_{2}\{\alpha\mu_{1}\}}\Big]\xi^{\lambda}\nonumber\\
&=&\Big[\partial_{\mu_{1}}B_{\lambda}^{[\alpha\mu_{1}]}-\partial_{\mu_{1}}B_{\lambda}^{\{\alpha\mu_{1}\}}\Big]\xi^{\lambda}
\nonumber\\
&=&\big(-\partial_{\mu_{1}}B^{\mu_{1}\alpha}\big)\xi^{\lambda}.\label{B213a}
\end{eqnarray}
Passage from the second equality to the third is due to (\ref{A107f}) with $s=2$.
The term with first-order derivatives of $\xi$ is 
\begin{eqnarray}
X^{\alpha}_{1}&=&\Big[{1\over 2}C_{\lambda}^{[\alpha\mu_{1}]}-{1\over 3}\partial_{\mu_{2}}C_{\lambda}^{[\alpha\mu_{1}]\mu_{2}}
+{2\over 3}\partial_{\mu_{2}}C_{\lambda}^{[\alpha\mu_{2}]\mu_{1}}\Big]\partial_{\mu_{1}}\xi^{\lambda}\nonumber\\
&=&\Big[B_{\lambda}^{[\alpha\mu_{1}]}-\partial_{\mu_{2}}B_{\lambda}^{\mu_{2}\{\alpha\mu_{1}\}}\Big]\partial_{\mu_{1}}\xi^{\lambda}\nonumber\\
&=&\Big[B_{\lambda}^{[\alpha\mu_{1}]}+B_{\lambda}^{\{\alpha\mu_{1}\}}\Big]\partial_{\mu_{1}}\xi^{\lambda}\nonumber\\
&=&B_{\lambda}^{\alpha\mu_{1}}\partial_{\mu_{1}}\xi^{\lambda}.\label{B213b}
\end{eqnarray}
Passage from the second to the third line uses  (\ref{A107f}) with $s=2$. 
The term with second-order derivatives of $\xi$ is
\begin{eqnarray}
X^{\alpha}_{2}&=&{2\over 3}\Big[B_{\lambda}^{\alpha\{\mu_{1}\mu_{2}\}}-B_{\lambda}^{\mu_{1}\{\alpha\mu_{2}\}}\Big]
\partial_{\mu_{1}}\partial_{\mu_{2}}\xi^{\lambda} \nonumber\\
&=&{1\over 3}\big[2B_{\lambda}^{\alpha\{\mu_{1}\mu_{2}\}}-B_{\lambda}^{\mu_{1}\{\alpha\mu_{2}\}}
-B_{\lambda}^{\mu_{2}\{\alpha\mu_{1}\}}\big]\partial_{\mu_{1}}\partial_{\mu_{2}}\xi^{\lambda}\nonumber\\
&=&B_{\lambda}^{\alpha\{\mu_{1}\mu_{2}\}}\partial_{\mu_{1}}\partial_{\mu_{2}}\xi^{\lambda}.\label{B213c}
\end{eqnarray}
where the second line results from symmetrizing in $\mu_{1}\mu_{2}$ and the third line follows from (\ref{B207}).
The sum of (\ref{B213a}), (\ref{B213b}),  and (\ref{B213c}) is
\begin{eqnarray}
\partial_{\mu_{1}}\Phi^{[\alpha\mu_{1}]}&=&-(\partial_{\mu_{1}}B^{\mu_{1}\alpha}\big)\xi^{\lambda}
+B_{\lambda}^{\alpha\mu_{1}}\partial_{\mu_{1}}\xi^{\lambda}\nonumber\\
&+&B_{\lambda}^{\alpha\{\mu_{1}\mu_{2}\}}\partial_{\mu_{1}}\partial_{\mu_{2}}\xi^{\lambda}
\end{eqnarray}
and this agrees with  (\ref{318}) or (\ref{A117}) with  $N=2$.

\subsection{The $N=3$ superpotential}
A derivative of the Ricci scalar $\nabla_{\mu}R$ or of the Ricci tensor $\nabla_{\mu}R_{\alpha\beta}$ is 
ultimately a
covariant derivative of the Riemann tensor. If  ${\cal L}$ contains such terms the Noether current 
contains third-order derivatives of $\xi$ and the superpotential must contain second-order derivatives:
\begin{eqnarray}
\Phi^{[\alpha\mu_{1}]}&=&W_{\lambda}^{[\alpha\mu_{1}]}\xi^{\lambda}
+W_{\lambda}^{[\alpha\mu_{1}]\mu_{2}}\partial_{\mu_{2}}\xi^{\lambda}\nonumber\\
&+&W_{\lambda}^{[\alpha\mu_{1}]\mu_{2}\mu_{3}}
\partial_{\mu_{2}}\partial_{\mu_{3}}\xi^{\lambda}.\label{B300}
\end{eqnarray}
The solution for the $W$'s is
\begin{eqnarray}
W_{\lambda}^{[\alpha\mu_{1}]\mu_{2}\mu_{3}}&=&{3\over 4}C_{\lambda}^{[\alpha\mu_{1}]\mu_{2}\mu_{3}}\label{B305a}\\
W_{\lambda}^{[\alpha\mu_{1}]\mu_{2}}&=&{2\over 3}C_{\lambda}^{[\alpha\mu_{1}]\mu_{2}}
-{1\over 2}\partial_{\mu_{3}}C_{\lambda}^{[\alpha\mu_{1}]\mu_{2}\mu_{3}}\label{B305b}\\
W_{\lambda}^{[\alpha\mu_{1}]}&=&{1\over 2}C_{\lambda}^{[\alpha\mu_{1}]}-{1\over 3}\partial_{\mu_{2}}
C_{\lambda}^{[\alpha\mu_{1}]\mu_{2}}\nonumber\\
&+&{1\over 4}\partial_{\mu_{2}}\partial_{\mu_{3}}C_{\lambda}^{[\alpha\mu_{1}]\mu_{2}\mu_{3}}
\Big].\label{B305c}
\end{eqnarray}
The proof that  $\partial_{\mu_{1}}\Phi^{[\alpha\mu_{1}]}$ agrees with  the current (\ref{A117}) for $N=3$
is a special case of the general proof in B.4.

\subsection{The superpotential for any $N$}
If ${\cal L}$ contains $N\!-\!2$ covariant derivatives of the Riemann tensor 
the conserved current  given in (\ref{A107f}) has $N$ derivatives of $\xi$ and the superpotential must have
$N-1$ derivatives of $\xi$:
\begin{equation}
\Phi^{[\alpha\mu_{1}]}=W_{\lambda}^{[\alpha\mu_{1}]}\xi^{\lambda}
+\sum_{s=2}^{N}W_{\lambda}^{[\alpha\mu_{1}]\mu_{2}..\mu_{s}}
\partial_{\mu_{2}}..\partial_{\mu_{s}}\xi^{\lambda}.\label{B401}
\end{equation}
The $W$ with $N+1$ superscripts is
\begin{equation}
W_{\lambda}^{[\alpha\mu_{1}]\mu_{2}..\mu_{N}}={N\over N+1}C_{\lambda}^{[\alpha\mu_{1}]\mu_{2}..\mu_{N}}.
\label{B405}\end{equation}
 $W$'s with fewer superscripts are given by the recursion relation
\begin{equation}
W_{\lambda}^{[\alpha\mu_{1}]\mu_{2}..\mu_{s}}={s\over s+1}\Big[C_{\lambda}^{[\alpha\mu_{1}]\mu_{2}..\mu_{s}}
-\partial_{\mu_{s+1}}W_{\lambda}^{[\alpha\mu_{1}]\mu_{2}..\mu_{s+1}}\Big]\label{B408}
\end{equation}
that relates the $W$ with $s+1$ superscripts to the derivative of the $W$ with $s+2$ superscripts. 
Using this once gives
\begin{eqnarray}
W_{\lambda}^{[\alpha\mu_{1}]\mu_{2}..\mu_{s}}&=&{s\over s\!+\!1}C_{\lambda}^{[\alpha\mu_{1}]..\mu_{s}}
-{s\over s\!+\! 2}\partial_{\mu_{s+1}}C_{\lambda}^{[\alpha\mu_{1}]\mu_{2}..\mu_{s+1}}\nonumber\\
&+&{s\over s+2}\partial_{\mu_{s+1}}\partial_{\mu_{s+2}}W_{\lambda}^{[\alpha\mu_{1}]\mu_{2}..\mu_{s+2}}.
\end{eqnarray}
Iteration leads to an explicit formula for $1\le s\le N$:
\begin{eqnarray}
&&W_{\lambda}^{[\alpha\mu_{1}]..\mu_{s}}={s\over s+1}C_{\lambda}^{[\alpha\mu_{1}]..\mu_{s}}\nonumber\\
&&\hskip0.5cm +\sum_{p=1}^{N-s}{s(-1)^{p}\over s+p+1}\partial_{\mu_{s+1}}..\partial_{\mu_{s+p}}
C_{\lambda}^{[\alpha\mu_{1}]..\mu_{s+p}}.\label{B411}
\end{eqnarray}

\subsection{Proof of the superpotential formula}
The derivative of the superpotential (\ref{B401}) is
\begin{eqnarray}
\partial_{\mu_{1}}\Phi^{[\alpha\mu_{1}]}&=&\big(\partial_{\mu_{1}}W_{\lambda}^{[\alpha\mu_{1}]}\big)\xi^{\lambda}\nonumber\\
&+&\sum_{s=1}^{N}Y_{\lambda}^{\alpha:\mu_{1}..\mu_{s}}\partial_{\mu_{1}}..\partial_{\mu_{s}}\xi^{\lambda},
\label{B415}\end{eqnarray}
where the coefficients are
\begin{equation}
Y_{\lambda}^{\alpha:\mu_{1}..\mu_{s}}=W_{\lambda}^{[\alpha\mu_{1}]\mu_{2}..\mu_{s}}
+\partial_{\mu_{s+1}}W_{\lambda}^{[\alpha\mu_{s+1}]\mu_{1}..\mu_{s}}.\label{B416}
\end{equation}
$Y$ is symmetric in $\mu_{2}...\mu_{s}$ but has no special property under $\alpha\mu_{1}$ exchange. 

\begin{widetext}

Substitution of (\ref{B411}) gives the explicit value 
\begin{eqnarray}
&&Y_{\lambda}^{\alpha:\mu_{1}..\mu_{s}}=
{s\over s+1}C_{\lambda}^{[\alpha\mu_{1}]..\mu_{s}}+\sum_{p=1}^{N-s}{(-1)^{p}\over s+p+1}
\partial_{\mu_{s+1}}..\partial_{\mu_{s+p}}\Big[ sC_{\lambda}^{[\alpha\mu_{1}]..\mu_{s+p}}
-(s+1)C_{\lambda}^{[\alpha\mu_{s+p}]\mu_{1}..\mu_{s+p-1}}\Big].\label{B418}
\end{eqnarray}
For (\ref{B415})  one needs the products of $Y$'s with $s$ derivatives of $\xi$ and it is convenient to organize them according to the number of derivatives $p$ acting on $C$:
\begin{eqnarray}
&&^{\alpha}Z_{s}^{0}={s\over s+1}C_{\lambda}^{[\alpha\mu_{1}]\mu_{2}..\mu_{s}}\partial_{\mu_{1}}..\partial_{\mu_{s}}\xi^{\lambda}\label{B420}\\
&&^{\alpha}Z_{s}^{p}={(-1)^{p}\over s+p+1}\partial_{\mu_{s+1}}..\partial_{\mu_{s+p}}
\Big[sC_{\lambda}^{[\alpha\mu_{1}]..\mu_{s+p}}-(s+1)C_{\lambda}^{[\alpha\mu_{s+p}]\mu_{1}..\mu_{s+p-1}}\Big]
\partial_{\mu_{1}}..\partial_{\mu_{s}}\xi^{\lambda}\label{B421}\\
&&Y_{\lambda}^{\alpha:\mu_{1}..\mu_{s}}\partial_{\mu_{1}}..\partial_{\mu_{s}}\xi^{\lambda}=
\sum_{p=0}^{N-s}\,^{\alpha}Z^{p}_{s}.\label{B419}
\end{eqnarray}
The divergence of the superpotential is the double sum
\begin{equation}
\partial_{\mu_{1}}\Phi^{[\alpha\mu_{1}]}=\big(\partial_{\mu_{1}}W_{\lambda}^{[\alpha\mu_{1}]}\big)\xi^{\lambda}
+\sum_{s=1}^{N}\sum_{p=0}^{N-s}\,^{\alpha}\!Z_{s}^{p}. \label{B423}
\end{equation}
In $^{\alpha}\!Z_{s}^{p}$ substitution of  the definition of the $C$'s gives
\begin{eqnarray}
\Big[sC_{\lambda}^{[\alpha\mu_{1}]..\mu_{s+p}}-(s+1)C_{\lambda}^{[\alpha\mu_{s+p}]\mu_{1}..\mu_{s+p-1}}\Big]
\partial_{\mu_{1}}..\partial_{\mu_{s}}\xi^{\lambda}&=&
\Big[-B_{\lambda}^{\alpha\{\mu_{1}..\mu_{s+p}\}}-sB_{\lambda}^{\mu_{1}\{\alpha\mu_{2}..\mu_{s+p}\}}\nonumber\\
&+&(s+1)B_{\lambda}^{\mu_{s+p}\{\alpha\mu_{1}..\mu_{s+p-1}\}}\Big]\partial_{\mu_{1}}..\partial_{\mu_{s}}\xi^{\lambda}.
\end{eqnarray}
The second term in the square brackets on the right can be symmetrized over $\mu_{1}..\mu_{s}$:
\begin{equation}
sB_{\lambda}^{\mu_{1}\{\alpha\mu_{2}..\mu_{s+p}\}}\partial_{\mu_{1}}..\partial_{\mu_{s}}\xi^{\lambda}
=\sum_{j=1}^{s}B_{\lambda}^{\mu_{j}\{\alpha..\mu_{s+p}\}^{\prime}}\partial_{\mu_{1}}..\partial_{\mu_{s}}\xi^{\lambda}
\end{equation}
where the prime indicates that $\mu_{j}$ is omitted from the enclosed list. 
This allows $^{\alpha}\!Z^{p}_{s}$ to be expressed as
\begin{equation}
^{\alpha}\!Z^{p}_{s}={(-1)^{p}\over s+p+1}\partial_{\mu_{s+1}}..\partial_{\mu_{s+p}}
\Big[-B_{\lambda}^{\alpha\{\mu_{1}..\mu_{s+p}\}}-\sum_{j=1}^{s}B_{\lambda}^{\mu_{j}\{\alpha\mu_{2}..\mu_{s+p}\}^{\prime}}\\
+(s+1)B_{\lambda}^{\mu_{s+p}\{\alpha\mu_{1}..\mu_{s+p-1}\}}\Big]\partial_{\mu_{1}}..\partial_{\mu_{s}}\xi^{\lambda}
.\label{B425}\end{equation}

\underbar{The largest $p$:} When $p=N-s$ all the $B$'s in (\ref{B425})  have $N+1$ superscripts and  satisfy the constraint
\begin{equation}
0=B_{\lambda}^{\{\alpha\mu_{1}..\mu_{N}\}}={1\over N+1}
\Big[B_{\lambda}^{\alpha\{\mu_{1}..\mu_{N}\}}+\sum_{j=1}^{N}B_{\lambda}^{\mu_{j}\{\alpha...\}^{\prime}}\Big].\label{B427}\end{equation}
Consequently
\begin{equation}
^{\alpha}\!Z^{N-s}_{p}={(-1)^{N-s}\over N+1}\partial_{\mu_{s+1}}..\partial_{\mu_{N}}\Big[\sum_{j=s+1}^{N}
B_{\lambda}^{\mu_{j}\{\alpha..\}^{\prime}}+(s+1)B_{\lambda}^{\mu_{N}\{\alpha\mu_{1}..\mu_{N-1}\}}\Big]
\partial_{\mu_{1}}..\partial_{\mu_{s}}\xi^{\lambda}.
\end{equation}
In the sum on $j$, the derivatives impose symmetry in $\mu_{s+1}..\mu_{N}$ which implies
\begin{equation}
\partial_{\mu_{s+1}}..\partial_{\mu_{N}}\sum_{j=s+1}^{N}B_{\lambda}^{\mu_{j}\{\alpha..\}^{\prime}}
=\partial_{\mu_{s+1}}..\partial_{\mu_{N}}(N-s)B_{\lambda}^{\mu_{N}\{\alpha\mu_{1}..\mu_{N-1}\}}
\end{equation}
and leads to
\begin{equation}
^{\alpha}\!Z^{N-s}_{s}={(-1)^{N-s}\over N+1}\partial_{\mu_{s+1}}..\partial_{\mu_{N}}(N+1)B_{\lambda}^{\mu_{N}\{\alpha\mu_{1}
..\mu_{N-1}\}}=(-1)^{N-s-1}\partial_{\mu_{s+1}}..\partial_{\mu_{N-1}}B_{\lambda}^{\{\alpha\mu_{1}..\mu_{N-1}\}}
.\end{equation}
Before proceeding it is useful to rewrite this.
The $B$'s with fewer than $N+1$ superscripts do not satisfy a constraint like (\ref{B427}). Fortunately all one needs is to express
the $B$ with $N$ symmetrized superscripts as a sum over $N$ terms each of which has  $N-1$ symmetrized superscripts
\begin{equation}^{\alpha}\!Z_{s}^{N-s}={(-1)^{N-s-1}\over N}\partial_{\mu_{s+1}}..\partial_{\mu_{N-1}}
\Big[B_{\lambda}^{\alpha\{\mu_{1}..\mu_{N-1}\}}
+\sum_{j=1}^{N-1}B_{\lambda}^{\mu_{j}\{\alpha..\}^{\prime}}\Big].\label{B429}
\end{equation}
This will be used in the following.

\underbar{The next largest $p$:} For the case $p=N\!-\!s\!-\!1$,  Eq. (\ref{B425}) gives
\begin{equation}
^{\alpha}\!Z^{N-s-1}_{s}={(-1)^{N-s-1}\over N}\partial_{\mu_{s+1}}..\partial_{\mu_{N-1}}\Big[-B_{\lambda}^{\alpha\{\mu_{1}..\mu_{N-1}\}}
-\sum_{j=1}^{s}B_{\lambda}^{\mu_{j}\{\alpha..\}^{\prime}}+(s+1)B_{\lambda}^{\mu_{N-1}\{\alpha\mu_{1}..\mu_{N-2}\}}
\Big]\partial_{\mu_{1}}..\partial_{\mu_{s}}\xi^{\lambda}.
\end{equation}
The sum of this with (\ref{B429}) is
\begin{equation}
^{\alpha}\!Z_{s}^{N-s-1}+\,^{\alpha}\!Z_{s}^{N-s}={(-1)^{N-s-1}\over N}\partial_{\mu_{s+1}}..\partial_{\mu_{N-1}}\Big[
-\sum_{j=s+1}^{N-1}B_{\lambda}^{\mu_{j}\{\alpha..\}^{\prime}}+(s+1)B_{\lambda}^{\mu_{N-1}\{\alpha\mu_{1}..\mu_{N-2}\}}\Big]
\partial_{\mu_{1}}..\partial_{\mu_{s}}\xi^{\lambda}.
\end{equation}
In the sum on $j$ the derivatives impose symmetry in $\mu_{s+1}\dots\mu_{N-1}$ and  allows
the sum to be replaced by a single term  $(N\!-\!s\!-\!1)B_{\lambda}^{\mu_{N-1}\{\alpha\mu_{1}..\mu_{N-2}\}}$.
When combined with the second term the result is  $NB_{\lambda}^{\mu_{N-1}\{\alpha\mu_{1}..\mu_{N-2}\}}$
and gives
\begin{eqnarray}
^{\alpha}\!Z_{s}^{N-s-1}+\,^{\alpha}\!Z_{s}^{N-s}&=&(-1)^{N-s-1}\partial_{\mu_{s+1}}..\partial_{\mu_{N-1}}B_{\lambda}^{\mu_{N-1}\{\alpha\mu_{1}..\mu_{N-2}\}}
\partial_{\mu_{1}}..\partial_{\mu_{s}}\xi^{\lambda}\nonumber\\
&=&(-1)^{N-s-2}\partial_{\mu_{s+1}}..\partial_{\mu_{N-2}}B_{\lambda}^{\{\alpha\mu_{1}..\mu_{N-2}\}}
\partial_{\mu_{1}}..\partial_{\mu_{s}}\xi^{\lambda}\label{B431}
\end{eqnarray}
after using (\ref{A107f}). 

\end{widetext}

\underbar{The sum on $p$:} The pattern set in (\ref{B431}) continues and leads to the sum
\begin{eqnarray}
\sum_{p=1}^{N-s}\,^{\alpha}Z_{s}^{p}&=&-\partial_{\mu_{s+1}}B_{\lambda}^{\mu_{s+1}\{\alpha\mu_{1}..\mu_{s}\}}\partial_{\mu_{1}}..\partial_{\mu_{s}}\xi^{\lambda}\nonumber\\
&=&B_{\lambda}^{\{\alpha\mu_{1}..\mu_{s}\}}\partial_{\mu_{1}}..\partial_{\mu_{s}}\xi^{\lambda}\\
&=&\hskip-0.1cm {1\over s+1}\Big[B_{\lambda}^{\alpha\{\mu_{1}..\mu_{s}\}}+sB^{\mu_{1}\{\alpha\mu_{2}..\mu_{s}\}}\Big]
\partial_{\mu_{1}}..\partial_{\mu_{s}}\xi^{\lambda}.\nonumber
\end{eqnarray}
From Eq. (\ref{B420}) the $p=0$ term is
\begin{equation}
^{\alpha}\!Z^{0}_{s}={s\over s+1}\Big[B_{\lambda}^{\alpha\{\mu_{1}..\mu_{s}\}}-B_{\lambda}^{\mu_{1}\{\alpha\mu_{2}..\mu_{s}\}}
\Big]\partial_{\mu_{1}}..\partial_{\mu_{s}}\xi^{\lambda}
\end{equation}
and so the full sum on $p$ is
\begin{equation}
\sum_{p=0}^{N-s}\,^{\alpha}\!Z_{s}^{p}=B_{\lambda}^{\alpha\{\mu_{1}..\mu_{s}\}}\partial_{\mu_{1}}..\partial_{\mu_{s}}\xi^{\lambda}.
\label{B433}\end{equation}

\underbar{The complete sum:}
The divergence of the superpotential as given by (\ref{B423}) is now
\begin{displaymath}
\partial_{\mu_{1}}\Phi^{[\alpha\mu_{1}]}=\big(\partial_{\mu_{1}}W_{\lambda}^{[\alpha\mu_{1}]}\big)\xi^{\lambda}
+\sum_{s=1}^{N}B_{\lambda}^{\alpha\{\mu_{1}..\mu_{s}\}}\partial_{\mu_{1}}..\partial_{\mu_{s}}\xi^{\lambda}.
\end{displaymath}
One needs $W_{\lambda}^{[\alpha\mu_{1}]}$ from (\ref{B411}) 
\begin{eqnarray}
\partial_{\mu_{1}}W_{\lambda}^{[\alpha\mu_{1}]}&=&{1\over 2}\partial_{\mu_{1}}C_{\lambda}^{[\alpha\mu_{1}]}\\
&+&\sum_{p=1}^{N-1}{(-1)^{p}\over p+2}\partial_{\mu_{1}}..\partial_{\mu_{p+1}}
C_{\lambda}^{[\alpha\mu_{1}]\mu_{2}..\mu_{p+1}}.\nonumber
\end{eqnarray}
The first term is $\partial_{\mu_{1}}B_{\lambda}^{[\alpha\mu_{1}]}$ as noted in (\ref{B103}). Evaluating the  sum over $p$
 requires the same approach as above: compute the term with the highest value of $p$; add to that the term with the
 next highest value of $p$; continue adding terms with successively smaller values of $p$.
 The sum from $p=1$ to $p=N-1$ gives $-\partial_{\mu_{1}}B_{\lambda}^{\{\alpha\mu_{1}\}}$ and so 
\begin{eqnarray}
\partial_{\mu_{1}}W_{\lambda}^{[\alpha\mu_{1}]}&=&\partial_{\mu_{1}}B_{\lambda}^{[\alpha\mu_{1}]}
-\partial_{\mu_{1}}B_{\lambda}^{\{\alpha\mu_{1}\}}\nonumber\\
&=&-\partial_{\mu_{1}}B_{\lambda}^{\mu_{1}\alpha}.
\end{eqnarray}
The final result is
\begin{equation}
\partial_{\mu_{1}}\Phi^{[\alpha\mu_{1}]}=-\big(\partial_{\mu_{1}}B_{\lambda}^{\mu_{1}\alpha}\big)\xi^{\lambda}
+\sum_{s=1}^{N}B_{\lambda}^{\alpha\{\mu_{1}..\mu_{s}\}}\partial_{\mu_{1}}..\partial_{\mu_{s}}\xi^{\lambda}.
\label{B440}\end{equation}
This is the current ${\cal J}^{\alpha}_{G2}$ given in (\ref{A117}) which proves that  the superpotential given in (\ref{B401}) with coefficients (\ref{B411})  is correct.

\underbar{Covariant form:} To express the generalized Einstein tensor or the superpotential or
various other quantities covariantly requires introducing for each covariant derivative of order $s$ of the Riemann 
tensor in ${\cal L}_{G}$ a tensor
\begin{equation}
\sqrt{g}\,M^{\{\mu_{1}..\mu_{s}\}\alpha_{1}..\alpha_{4}}=
{\partial{\cal L}_{G}\over\partial(\nabla_{\{\mu_{1}}..\nabla_{\mu_{s}\}}R_{\alpha_{1}..\alpha_{4}}) }.
\end{equation}

\section{Covariant results for $N=2$ \label{N=2 Covariant}}

\subsection{Tensor form for the $N=2$ Noether current \label{N=2 Current}}
This section will prove the result (\ref{341}) starting from Eq. (\ref{304}) for
the Noether current:
\begin{eqnarray}
{\cal J}^{\alpha}_{G}&=&\bigg[{\partial{\cal L}_{G}\over\partial(\partial_{\alpha}g_{\rho\sigma})}
-\partial_{\nu}{\partial{\cal L}_{G}\over\partial(\partial_{\alpha}\partial_{\nu}g_{\rho\sigma})}\bigg]\pounds_{\xi}g_{\rho\sigma}\nonumber\\
&&+{\partial{\cal L}_{G}\over\partial(\partial_{\alpha}\partial_{\nu}g_{\rho\sigma})}\partial_{\nu}\pounds_{\xi}g_{\rho\sigma}
-\xi^{\alpha}{\cal L}_{G}.\end{eqnarray}
The current is a sum of
 three functional derivatives  
\begin{equation}
{\cal J}^{\alpha}_{G}={\cal V}^{\alpha}_{1}+{\cal V}^{\alpha}_{2}+{\cal V}^{\alpha}_{3}-\xi^{\alpha}{\cal L}\label{D100}\end{equation}
which will now be calculated in terms of  the tensor $M^{\alpha\mu\nu\beta}$  defined in (\ref{338}) .

The first functional derivative gives
\begin{eqnarray}
{\cal V}^{\alpha}_{1}&=&{\partial{\cal L}\over\partial(\partial_{\alpha}g_{\mu\nu})}\pounds_{\xi}g_{\mu\nu}\nonumber\\
&=&\sqrt{g}\,\Big[2\Gamma^{\mu}_{\rho\omega}
M^{\rho\{\nu\alpha\}\omega}+2\Gamma^{\nu}_{\rho\omega}
M^{\rho\{\mu\alpha\}\omega}\nonumber\\
&&-2\Gamma^{\alpha}_{\rho\omega}M^{\rho\{\mu\nu\}\omega}\Big]\pounds_{\xi}g_{\mu\nu}\label{D110}
\end{eqnarray}
The second functional derivative  gives
\begin{eqnarray}
{\cal V}^{\alpha}_{2}&=&-\Big[\partial_{\beta}{\partial{\cal L}\over\partial(\partial_{\alpha}\partial_{\beta}g_{\mu\nu})}\Big]
\pounds_{\xi}g_{\mu\nu}\nonumber\\
&=&-\Big[\partial_{\beta}\Big(2\sqrt{g}\,M^{\alpha\{\mu\nu\}\beta}\Big)\Big]\pounds_{\xi}g_{\mu\nu}\nonumber\\
&=&2\sqrt{g}\,\Big[ -\nabla_{\beta}M^{\alpha\{\mu\nu\}\beta}
+\Gamma^{\alpha}_{\rho\omega}M^{\rho\{\mu\nu\}\omega}\nonumber\\
&+&\Gamma^{\mu}_{\rho\omega}M^{\alpha\{\rho\nu\}\omega}
+\Gamma^{\nu}_{\rho\omega}M^{\alpha\{\rho\mu\}\omega}\Big]\pounds_{\xi}g_{\mu\nu}.\label{D120}
\end{eqnarray}
 The third functional derivative gives
\begin{eqnarray}
{\cal V}^{\alpha}_{3}&=&{\partial{\cal L}\over\partial(\partial_{\alpha}\partial_{\beta}g_{\mu\nu})}\partial_{\beta}
\pounds_{\xi}g_{\mu\nu}\nonumber\\
&=&2\sqrt{g}\,M^{\alpha\{\mu\nu\}\beta}\partial_{\beta}
\pounds_{\xi}g_{\mu\nu}\nonumber\\
&=&2\sqrt{g}\,M^{\alpha\{\mu\nu\}\beta}\nabla_{\beta}\pounds_{\xi}g_{\mu\nu}\nonumber\\
&+&2\sqrt{g}\,\Big[\Gamma^{\mu}_{\rho\omega}M^{\alpha\{\rho\nu\}\omega}
+\Gamma^{\nu}_{\rho\omega}M^{\alpha\{\mu\rho\}\omega}\Big]\pounds_{\xi}g_{\mu\nu}.\label{D130}
\end{eqnarray}
The sum of (\ref{D110}), (\ref{D120}), and (\ref{D130}) is
\begin{eqnarray}
&&{\cal V}^{\alpha}_{1}\!+\!{\cal V}^{\alpha}_{2}\!+\!{\cal V}^{\alpha}_{3}\nonumber\\
&&\hskip0.8cm =
2\sqrt{g}\,\Big[-\nabla_{\beta}M^{\alpha\{\mu\nu\}\beta}\pounds_{\xi}g_{\mu\nu}+\!M^{\alpha\{\mu\nu\}\beta}\nabla_{\beta}\pounds_{\xi}g_{\mu\nu}\Big]\nonumber\\
&&\hskip1cm +\sqrt{g}\,\Gamma^{\mu}_{\rho\omega}\Big\{2M^{\rho\{\alpha\nu\}\omega} +4M^{\alpha\{\rho\nu\}\omega}\Big\}\pounds_{\xi}g_{\mu\nu}\label{D180}\\
&&\hskip1cm +\sqrt{g}\,\Gamma^{\nu}_{\rho\omega}\Big\{2M^{\rho\{\alpha\mu\}\omega}+4M^{\alpha\{\rho\mu\}\omega}
\Big\}\pounds_{\xi}g_{\mu\nu}.\nonumber
\end{eqnarray}
By using the relations (\ref{340}) the first quantity in braces can be simplified to
\begin{equation}
\Gamma^{\mu}_{\rho\omega}\Big\{M^{\alpha\omega\rho\nu}-M^{\alpha\rho\omega\nu}+2M^{\alpha\nu\rho\omega}\Big\}.\label{D190}\end{equation}
This vanishes because the quantity in braces is antisymmetric under $\rho\omega$ interchange.
The second quantity in braces in  (\ref{D180}) vanishes similarly.

The total Noether current is therefore
\begin{eqnarray}
{\cal J}^{\alpha}_{G}&=&-2\sqrt{g}\,(\nabla_{\beta}M^{\alpha\{\mu\nu\} \beta})\pounds_{\xi}g_{\mu\nu}\nonumber\\
&&+2\sqrt{g}\,M^{\alpha\{\mu\nu\}\beta}\nabla_{\beta}\pounds_{\xi}g_{\mu\nu}\\
&&-\xi^{\alpha}\sqrt{g}\,L_{G}\nonumber
\end{eqnarray}
and is explicitly a vector density. 
Some relabeling in the second line converts this to
\begin{eqnarray}
{\cal J}^{\alpha}_{G}&=&-2\sqrt{g}(\nabla_{\beta}{\rm M}^{\alpha\{\mu\nu\}\beta})\pounds_{\xi}g_{\mu\nu}\nonumber\\
&&+4\sqrt{g}\,{\rm M}^{\alpha\{\beta\nu\}\mu}\nabla_{\mu}\nabla_{\nu}\xi_{\beta}\label{D200}\\
&&-\xi^{\alpha}\sqrt{g}\,L_{G}.\nonumber\end{eqnarray}
which is the result used in (\ref{341}).

\subsection{Tensor form for the $N=2$ superpotential\label{N=2 Superpotential}}
The superpotential given in Eq.  (\ref{350}) 
\begin{equation}
\Phi^{[\alpha\mu]}=\Big[B^{[\alpha\mu]}_{\lambda}-{1\over 3}\partial_{\nu}C^{[\alpha\mu]\nu}_{\lambda}\Big]\xi^{\lambda}+{2\over 3}C^{[\alpha\mu]\nu}_{\lambda}\partial_{\nu}\xi^{\lambda}
\label{E100}\end{equation}
was proven correct in Appendix B.1. 
This section will express the superpotential in a manifestly
covariant form. The first step is to expand the covariant  expression for the current (\ref{D200})
in a derivative expansion:
\begin{equation}
{\cal J}_{G}^{\alpha}=A^{\alpha}_{\lambda}\xi^{\lambda}+B^{\alpha\mu}_{\lambda}\partial_{\mu}\xi^{\lambda}
+B^{\alpha\{\mu\nu\}}_{\lambda}\partial_{\mu}\partial_{\nu}\xi^{\lambda}.\end{equation}
Only the antisymmetric part $B^{[\alpha\mu]}_{\lambda}=(B^{\alpha\mu}_{\lambda}
-B^{\mu\alpha}_{\lambda})/2$ is needed:
\begin{eqnarray}
B^{[\alpha\mu]}_{\lambda}&=&3\sqrt{g}(\nabla_{\nu}M^{\alpha\mu\nu\beta})g_{\beta\lambda}
-3\Gamma_{\beta,\rho\lambda}{\cal M}^{\alpha\mu\rho\beta}\nonumber\\
&&+\sqrt{g}\,(M^{\alpha\rho\omega\nu}\Gamma^{\mu}_{\rho\omega}
-M^{\mu\rho\omega\nu}\Gamma^{\alpha}_{\rho\omega})g_{\nu\lambda}.\label{E120}
\end{eqnarray}
Also needed is
\begin{eqnarray} 
C^{[\alpha\mu]\nu}_{\lambda}
&=&B^{\alpha\{\mu\nu\}}_{\lambda}-B^{\mu\{\alpha\nu\}}_{\lambda}\nonumber\\
&=&-3\sqrt{g}\,M^{\alpha\mu\nu\beta}g_{\beta\lambda}.\label{E125}\end{eqnarray}
The  derivative of this is
\begin{eqnarray}
-{1\over 3}\partial_{\nu}C^{[\alpha\mu]\nu}_{\lambda}
&=&\sqrt{g}(\nabla_{\nu}M^{\alpha\mu\nu\beta})g_{\beta\lambda}+\sqrt{g}\,\Gamma_{\beta,\rho\lambda}M^{\alpha\mu\rho\beta}\nonumber\\
&+&\sqrt{g}\,(\Gamma^{\alpha}_{\rho\omega}M^{\mu\rho\omega\beta}
-\Gamma^{\mu}_{\rho\omega}M^{\alpha\rho\omega\beta})g_{\beta\lambda}.\label{E130}
\end{eqnarray}
The sum of (\ref{E120}) and (\ref{E130}) when multiplied by $\xi^{\lambda}$ is
\begin{eqnarray}
&&\Big[B^{[\alpha\mu]}_{\lambda}-{1\over 3}\partial_{\nu}C^{[\alpha\mu]\nu}_{\lambda}\Big]\xi^{\lambda}\label{E140}\\
&&\hskip1.2cm=4\sqrt{g}(\nabla_{\nu}M^{\alpha\mu\nu\beta})\xi_{\beta}
-2\sqrt{g}\,M^{\alpha\mu\rho\beta}\Gamma_{\beta,\rho\lambda}\xi^{\lambda}.\nonumber
\end{eqnarray}
Equation (\ref{E100})  requires the combination
\begin{equation}
{2\over 3}C^{[\alpha\mu]\nu}_{\lambda}\partial_{\nu}\xi^{\lambda}=
2\sqrt{g}\,\Big[-M^{\alpha\mu\nu\beta}\nabla_{\nu}\xi_{\beta}
+M^{\alpha\mu\rho\beta}\Gamma_{\beta,\rho\lambda}\xi^{\lambda}\Big].\label{E150}
\end{equation}
The sum of (\ref{E140}) and (\ref{E150}) gives the tensor result
\begin{equation}
\Phi^{[\alpha\mu]}=4\sqrt{g}(\nabla_{\nu}M^{\alpha\mu\nu\beta})\xi_{\beta} -2\sqrt{g}\,M^{\alpha\mu\nu\beta}\nabla_{\nu}\xi_{\beta}\label{E160}
\end{equation}
which is used in Eq. (\ref{343a}) and subsequently.

\end{appendix}


\begin{thebibliography}{}

\bibitem{Noether} E. Noether, Invariante Variationsprobleme,
{\it Nach. d. K\"onig. Gesellsch. d. Wiss. zu G\"ottingen, Math-phys. Kasse}, 235-257 (1918).
(English translation by M.A. Tavel  arXiv:physics/0503066.)

\bibitem{Utiyama} R. Utiyama, New Energy and Energy-Tensor of a Gravitational Field,
Prog. Theor. Phys. {\bf 72}, 83 (1984).

\bibitem{DeHaro} S. DeHaro, Noether's Theorems and Energy in General Relativity, in {\it The Philosophy and Physics
of Noether's Theorems}, p. 197-256, ed. J. Read and N. Teh (Cambridge Univ. Press, Cambridge, U.K. 2022).
arXiv:2103.17160.

\bibitem{McLeod} R.J. McLeod, A Brief Review of Noether's Theorems \\ and their Applications to General Relativity,
 arXiv:2106.04393.

\bibitem{Ilin1} R.  Ilin and S. Paston, Energy-momentum Pseudotensor and Superpotential for Generally Covariant Theories of Gravity of General Form, Universe {\bf 6}, 173 (2020).

\bibitem{Ilin2} R.V. Ilin, Noether currents in theories with higher derivatives and theories with differential field transformation in action (DFTA), Eur. Phys. J. Plus, {\bf 137}, 1144 (2022).

\bibitem{Aoki} S. Aoki and T. Onogi, Conserved non-Noether charge in general relativity:
Physical definition versus Noether's second theorem, Int. Jour. Mod. Phys. A, {\bf 37}, 2250129 (2022).

\bibitem{eSa1} N.B. e S{\'a}, M.A.S. Pinto, and T. Trindale,
Covariant conservation laws, local invariance and Noether's second theorem,
Eur. Phys. J. C {\bf 86}, 612 (2026).

\bibitem{eSa2} N.B. e S{\'a}, Improper currents in theories with local invariance,
Ann. Phys. {\bf 485}, 170331 (2026).

\bibitem{Freese} A. Freese, Reflections on Noether's second theorem and the energy-momentum 
tensor, Phys. Rev. D {\bf 113}, 016011 (2026).

\bibitem{Einstein} A. Einstein, Die Grundlage der allgemeinen Relativit\"atstheorie, Ann. der Physik, {\bf 49}, 769 (1916).
English translation in H.A. Lorentz, A. Einstein, H. Minkowski, and H. Weyl,    {\it The Principle of Relativity}
(Dover, Toronto, Canada, 1952), p. 149.


\bibitem{Moller} C. M{\o}ller, On the localization of the energy of a physical system in the general theory
of relativity, Ann. Phys. (N.Y.) {\bf 4}, 347 (1958).

\bibitem{Dirac} P.A.M. Dirac, {\it General Theory of Relativity} (Princeton Univ. Press, Princeton N.J., 1996), \S 31 - \S32.

\bibitem{Landau}  L.D. Landau and E.M. Lifshitz, {\it The Classical Theory of Fields} (Pergamon Press, Oxford Eng. 1975),
\S 96.

\bibitem{Komar1} A. Komar, Covariant Conservation Laws in General Relativity, Phys. Rev.  {\bf 113},
934 (1959).

\bibitem{Komar2} A. Komar, Asymptotic Covariant Conservation Laws for Gravitational Radiation,
Phys. Rev. {\bf 127}, 1411 (1962).

\bibitem{Komar3} A. Komar, Positive-Definite Energy Density and Global Consequences for General Relativity,
Phys. Rev.  {\bf 129}, 1873 (1962).

\bibitem{Townsend} P.K. Townsend, Black Holes, arXiv:gr-qc/9707012.

\bibitem{Capozziello1} S. Capozziello and M. De Laurentis, Extended Theories of Gravity, Phys. Reps.
{\bf 509}, 167 (2011).

\bibitem{Clifton} T. Clifton, P.G. Ferreira, A. Padilla, and C. Skordis, Modified Gravity and Cosmology,
Phys. Reps. {\bf 513}, 1 (2012).


\bibitem{Odintsov1}  S. Nojiri and S.D. Odintsov, Unified cosmic history in modified gravity: From $F(R)$ theory
to Lorentz non-invariant models, Phys. Rep. {\bf 505}, 59 (2011).


\bibitem{Odintsov2}   S. Nojiri, S.D. Odintsov, and V.K. Oikonomou, Modified gravity theories on a nutshell: Inflation, bounce and late-time evolution, Phys. Rep. {\bf 692}  1 (2017).  



\bibitem{Capozziello2} S. Capozziello, V.F. Cardone, and V. Salzano, Cosmography of f(R) gravity,
Phys. Rev. D {\bf 78}, 064504 (2008).

\bibitem{DeFelice1} A. De Felice and S. Tsujikawa, f(R) Theories, Living Rev. Relativity, {\bf 13} 3 (2010).

\bibitem{Sotiriou} T.P. Sotiriou and V. Faraoni, f(R) Theories of Gravity, Rev. Mod. Phys. {\bf 82}, 451 (2010).

\bibitem{Kumar} S. Kumar, R.C. Nunes, S. Pan, and P. Yadav, New late-time constraints on f(R) gravity,
Physics of the Dark Universe, {\bf 42}, 101281 (2023).

\bibitem{Montani} G. Montani, L.A. Escamilla, N. Carlevaro, and E. Di Vanetino, Decay of f(R) quintessence into
dark matter: Mitigating the Hubble tension?, Phys. Rev. D {\bf 113}, 023507 (2026).


\bibitem{Alvarez-Gaume} L. Alvarez-Gaume, A. Kehagias, C. Dounnas, D. L\"ust, and A. Riotto,
Aspects of quadratic gravity, Fortschr. Phys. {\bf 64}, 176. (2016).

\bibitem{Holdom} B. Holdom and J. Ren, Quadratic gravity: From weak to strong,
Int. Jour. Mod. Phys. D, {\bf 25}, 1643004 (2016).

\bibitem{Salvio} A. Salvio, Quadratic Gravity, Front. Phys. {\bf 6:77} (2018).

\bibitem{Daas} J. Daas, K.Kuijpers, F. Saueressig, M.F. Wondrak, and H. Falcke, Probing
quadratic gravity with the Event Horizon Telescope, A \& A {\bf 673}, A53 (2023).

\bibitem{Donoghue}  J.F. Donoghue and G. Menezes, On quadratic gravity, Il Nuo. Cim. {\bf 45} C,  26 (2022).

\bibitem{Brito} G.P. de Brito, Quadratic gravity in analogy to quantum chromodynamics: Light fermions in its landscape, Phys. Rev. D {\bf 109}, 086005. (2024).

\bibitem{Buccio}  D. Buccio, J.F. Donoghue, G. Menezes, and R. Percacci, Physical Running of Couplings 
in Quadratic Gravity, Phys. Rev. Lett. {\bf 133}, 021604 (2024).

\bibitem{Kuntz} J. Kuntz, Unitarity through PT symmetry in quantum quadratic gravity,
Class. Quantum Grav. {\bf 42} 175003. (2025).


\bibitem{DeFelice2} A. De Felice and S. Tsujikawa, Solar system constraints on $f(G)$ models,
Phys. Rev. D {\bf 80}, 063516 (2009).

\bibitem{Glavan} D. Glavan and C. Lin, Einstein-Gauss-Bonnet gravity in four-dimensional spacetime,
Phys. Rev. Lett. {\bf 124}, 081301 (2020).

\bibitem{Fernandes} P.G.S. Fernandes, P. Carrilho, T. Clifton, and D.J. Mulryne,
The 4D Einstein-Gauss-Bonnet theory of gravity: a review, Class. Quantum Grav. {\bf 39},
063001 (2022). 

\bibitem{Ortin}    T. Ort\'in, Komar integrals for theories of higher order in the Riemann curvature
and black-hole chemistry, JHEP {\bf 08}, 023 (2021).  

\bibitem{Goldberg} J.N. Goldberg, Conservation Laws in General Relativity, Phys. Rev. {\bf 111},  315 (1958).

\bibitem{Bergmann} P. G. Bergmann, Conservation Laws in General Relativity as the Generators of Coordinate Transformations, Phys. Rev. {\bf 112}, 287 (1958).

\bibitem{Wald} V. Iyer and R.M. Wald, Comparison of the Noether charge and Euclidean methods for
computing the entropy of stationary black holes, Phys. Rev. D {\bf 52}, 4430 (1995).

\bibitem{Biswas} T. Biswas, E. Gerwick, T. Koivisto, and A. Maxumdar, Towards Singularity-and Ghost-Free Theories of Gravity, Phys. Rev. Lett. {\bf 108}, 031101 (2912).



\end{thebibliography}
\end{document}